\documentclass[sigconf]{acmart}

\usepackage{tikz}
\usetikzlibrary{calc}
\usepackage{url}
\usepackage{mdframed}
\usepackage{algpseudocode}
\usepackage{pgfplots}
\usepackage{pgfplotstable}
\usepackage{graphicx}
\usepackage{subcaption}
\usepackage{caption}
\usetikzlibrary{calc}
\usetikzlibrary{patterns}
\usetikzlibrary{shapes,snakes}
\usetikzlibrary{shapes.multipart, arrows}
\usetikzlibrary{decorations.pathreplacing}
\usepackage{multirow}
\usepackage{balance}
\usepackage{algorithm,algpseudocode}
\usepackage{amsmath} 
\usepackage{xcolor,colortbl}
\usepackage{enumitem}
\usepackage{tabularx}
\usepackage{pdflscape}
\usepackage{multicol}

\makeatletter
\def\@copyrightspace{\relax}
\makeatother

\def\poet{\textproc{PoET}}
\def\spoet{\textproc{PoET+}}

\setcopyright{none} 

\makeatletter
\renewcommand\footnotetextcopyrightpermission[1]{}

\author{Hung Dang, Tien Tuan Anh Dinh, Dumitrel Loghin \\ Ee-Chien Chang, Qian Lin, Beng Chin Ooi}
\affiliation{
  \institution{National University of Singapore}
}
\email{{hungdang,dinhtta,dumitrel,changec,linqian,ooibc}@comp.nus.edu.sg}

\makeatother

\begin{document}  \sloppy
\definecolor{palatinatepurple}{rgb}{0.41, 0.16, 0.38}
\definecolor{prussianblue}{rgb}{0.0, 0.19, 0.33}
\definecolor{green}{rgb}{0.0, 0.46, 0.25}
\definecolor{brown}{rgb}{0.73, 0.4, 0.16}
\definecolor{blond}{rgb}{0.98, 0.94, 0.75}
\definecolor{airforceblue}{rgb}{0.36, 0.54, 0.66}
\definecolor{amaranth}{rgb}{0.9, 0.17, 0.31}
\definecolor{asparagus}{rgb}{0.53, 0.66, 0.42}
\definecolor{bronze}{rgb}{0.8, 0.5, 0.2}
\definecolor{cardinal}{rgb}{0.77, 0.12, 0.23}
\definecolor{grad_0}{RGB}{235, 130, 128} 
\definecolor{grad_1}{RGB}{240, 185, 147} 
\definecolor{grad_2}{RGB}{245, 226, 150} 
\definecolor{grad_3}{RGB}{225, 235, 185} 
\definecolor{grad_4}{RGB}{220, 245, 215} 

\newpage

\title{Towards Scaling Blockchain Systems via Sharding}

\maketitle

\algnewcommand{\IfThenElse}[3]{
  \State \algorithmicif\ #1\ \algorithmicthen\ #2\ \algorithmicelse\ #3}
\section*{Abstract}
Existing blockchain systems scale poorly because of their distributed consensus protocols. Current attempts
at improving blockchain scalability are limited to cryptocurrency. 
Scaling blockchain systems under general workloads (i.e., non-cryptocurrency applications)
remains an open question.  

In this work, we take a principled approach to apply sharding, which is a well-studied and proven technique to
scale out databases, to blockchain systems in order to improve their transaction throughput at scale.  This is
challenging, however, due to the fundamental difference in failure models between databases and blockchain.
To achieve our goal, we first enhance the performance of Byzantine consensus protocols, by doing so we improve
individual shards' throughput.  Next, we design an efficient shard formation protocol that leverages a trusted
random beacon to securely assign nodes into shards. We rely on trusted hardware, namely Intel SGX, to achieve
high performance for both consensus and shard formation protocol. Third, we design a general distributed
transaction protocol that ensures safety and liveness even when transaction coordinators are malicious.
Finally, we conduct an extensive evaluation of our design both on a local
cluster and on Google Cloud Platform.
The results show that our consensus and shard formation protocols outperform state-of-the-art solutions at
scale. More importantly, our sharded blockchain reaches a high throughput that can handle Visa-level
workloads, and is the largest ever reported in a realistic environment. 

\section{Introduction}
\label{sec:intro}
Blockchain systems offer data transparency, integrity and immutability in a decentralized and potentially hostile
environment. These strong security guarantees come at a dear cost to scalability, for blockchain
systems have to rely on distributed consensus protocols which have been shown to scale poorly, both in terms of
number of transactions per second (tps) and number of nodes~\cite{blockbench}.

A number of works have attempted to scale consensus protocols, the ultimate goal being able to handle average
workloads of centralized systems such as Visa. One scaling approach is 
to exploit trusted hardware~\cite{mscoco, poet, hybster}. However, its effectiveness has not 
been demonstrated on data-intensive blockchain workloads. The second approach is to use 
sharding, a well-studied and proven technique to scale out databases, to divide the blockchain network into
smaller committees so as to reduce the overhead of consensus protocols. Examples of
sharded blockchains include Elastico~\cite{elastico},
OmniLedger~\cite{omniledger} and RapidChain~\cite{rapidchain}. These systems,
however, are limited to cryptocurrency applications in an open (or
permissionless) setting. Since they focus on a simple data model, namely the
unspent transaction output (UTXO) model, these approaches do not generalize to
applications beyond Bitcoin.

In this paper, we take a principled approach to extend {\em sharding} to permissioned blockchain systems.
Existing works on sharded blockchains target permissionless systems and focus on
security. Here, our focus is on performance. In particular, our goal is to design a blockchain system that can support a large network size
equivalent to that of major cryptocurrencies like Bitcoin~\cite{btc_origin} and Ethereum~\cite{eth_origin}. At
the same time, it achieves high transaction throughput that can handle the average  workloads of centralized
systems such as Visa, which is around $2,000-4,000$ transactions per second~\cite{bitcoinwiki_scalability}.
Finally, the system supports any blockchain application from domains such as finance~\cite{ripple}, supply
chain management~\cite{fr8} and healthcare~\cite{medilot}, not being limited to cryptocurrencies.

\begin{figure}
\begin{subfigure}{0.48\textwidth}
\centering
{\includegraphics[width=0.6\textwidth]{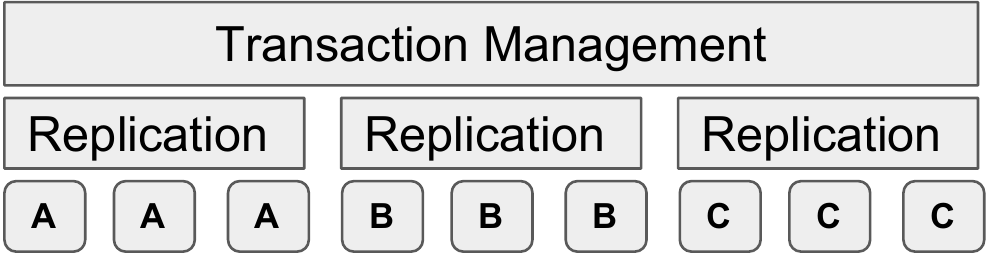}}
\caption{Distributed databases.}
\label{fig:shard_db}
\end{subfigure}
\par\bigskip
\begin{subfigure}{0.48\textwidth}
\centering
{\includegraphics[width=0.9\textwidth]{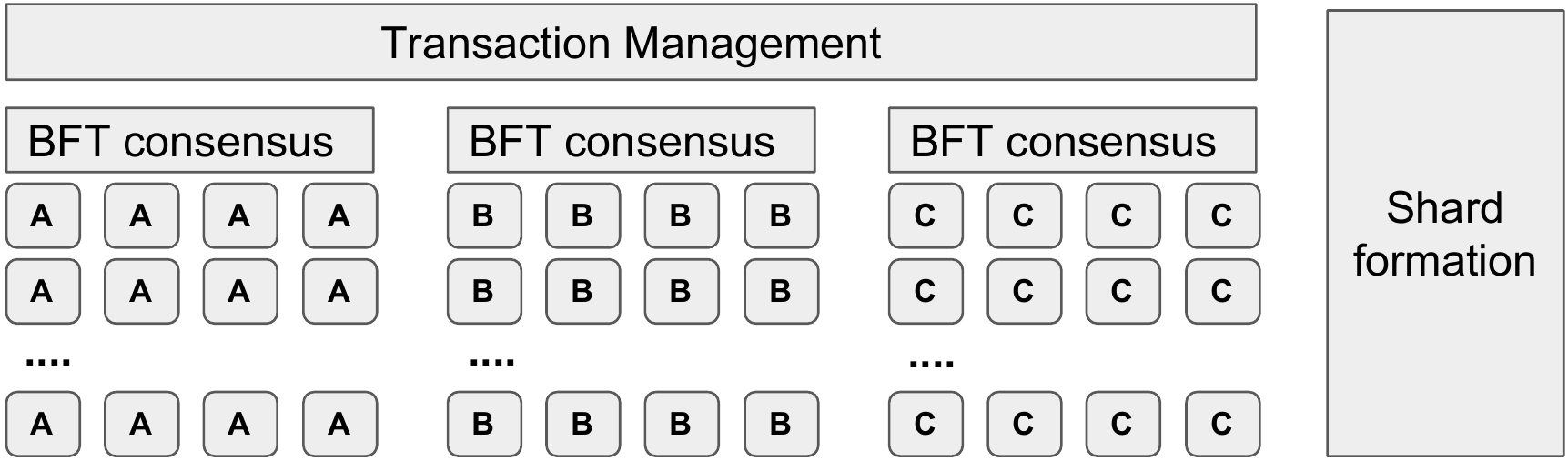}}
\caption{Sharded blockchains.}
\label{fig:shard_bc}
\end{subfigure}
\caption{Sharding protocols in traditional databases vs. blockchains.}
\label{fig:shard}
\end{figure}

Sharding protocols have been extensively studied in distributed database systems. A sharding protocol 
must ensure both  atomicity and isolation of transaction execution.
State-of-the-art protocols~\cite{google_spanner, janus, tapir, carousel} aim to improve performance for
distributed transactions in geo-replicated settings. However, they cannot be directly extended to
blockchain systems, due to a fundamental difference in the failure models that databases and blockchains consider. 
Traditional databases assume the crash-failure model, in which a faulty node
simply stops sending and responding to requests. On the other hand, blockchain systems operate in a more hostile environment, therefore
they assume a stronger failure model, namely Byzantine failure, to account for malicious attackers.
Figure~\ref{fig:shard} highlights the differences between distributed databases and sharded blockchains.  

The distinction in failure models leads to three challenges when applying database sharding techniques to
blockchains.  First, high-performance consensus protocols used in distributed databases~\cite{paxos, raft},
cannot be applied to blockchains. Instead, blockchains rely on Byzantine Fault Tolerance (BFT) consensus
protocols which have been shown to be a scalability bottleneck~\cite{blockbench}. Thus, the first
challenge is to scale BFT consensus protocols. 
Second, in a distributed database any node can belong to any shard, but a blockchain must assign nodes to
shards in a secure manner to ensure that no shard can be compromised by the attacker.  The second challenge,
therefore, is to achieve secure and efficient shard formation. Third, the distributed database relies on
highly available transaction coordinators to ensure atomicity and isolation, but coordinators in the
blockchain may be malicious.  Consequently, the third challenge is to enable secure distributed transactions
even when the coordinator is malicious.

We tackle the first challenge, that of improving BFT consensus protocols, by leveraging trusted execution
environment (TEE) (e.g., Intel SGX~\cite{sgx}) to eliminate equivocation in the Byzantine failure model.
Without equivocation,  existing BFT protocols can achieve higher fault tolerance with the same number of nodes
(i.e., a committee of $n$ nodes can tolerate up to $\frac{n-1}{2}$ non-equivocating Byzantine failures, as
opposed to $\frac{n-1}{3}$ failures in the original threat model~\cite{a2m, trinc, hybster}). We introduce
two major optimizations to the TEE-assisted BFT consensus protocol to lower its communications overhead,
which further improves the system throughput at scale.

We leverage the TEE to design an efficient and secure shard formation
protocol, addressing the second challenge. More specifically, we implement a
trusted randomness beacon inside the TEE to generate unbiased random values in a
distributed setting, which achieves significant speed-up over existing cryptographic
protocols~\cite{randhound, vrf}.
Furthermore, we exploit the increased fault tolerance of our TEE-assisted BFT
consensus protocol to reduce the shard size. In particular, to tolerate an
attacker who controls $25\%$ of the network, our sharding protocol requires
$80$-node committees as opposed to $600$-node committees used in related
works~\cite{elastico, omniledger}.

We tackle the final challenge, that of enabling distributed
transactions for general blockchain applications, by using two classic protocols, namely two-phase locking
(2PL) and two-phase commit (2PC). They ensure isolation and atomicity for cross-shard transactions.
Furthermore, they support any data model, as opposed to UTXO-optimized protocols that do not
generalize~\cite{rapidchain} beyond Bitcoin.  To prevent malicious transaction coordinators from causing infinite blocking as
in Omniledger~\cite{omniledger}, we design a protocol that runs the 2PC's coordination logic as a BFT
replicated state machine.

In summary, this paper makes the following contributions:

\begin{itemize}

\item To the best of our knowledge, we are the first to present a sharded
blockchain that supports workloads other than cryptocurrency and is able to
scale to few thousands of transactions per second.

\item We present our design of a sharded blockchain consisting of three key novelties: (i) optimizations that
improve the performance of the consensus protocol running within each individual shard, (ii) an efficient
shard formation protocol, and (iii) a secure distributed transaction protocol that handles cross-shard,
distributed transactions.

\item We conduct extensive, large-scale experiments to evaluate the performance
of our design. We run our experiments on a local cluster with $100$ nodes and on a
realistic setup consisting of over $1400$ Google Cloud Platform (GPC) nodes
distributed across $8$ regions. On GPC setup, we achieve a throughput
of over $3,000$ transactions per second which, to the best of our knowledge, is the largest ever reported in a realistic environment.

\end{itemize}

The remaining of this paper is structured as follows. 
Section~\ref{sec:background} provides background on database sharding
techniques, blockchain consensus protocols, and Intel SGX.
Section~\ref{sec:problem} describes the key challenges in extending database
sharding to blockchain.
Section~\ref{sec:existing_protocols} discusses how we improve the underlying
consensus protocol running in each individual shard.
Section~\ref{sec:shard_formation} discusses the committee formation protocol.
Section~\ref{sec:coordinating_shard} presents our  distributed transaction protocol.  Section~\ref{sec:eval}
reports the performance of our design.  In Section~\ref{sec:related_work} we discuss the related work, before
concluding in Section~\ref{sec:conclusion}.

\section{Preliminaries}
\label{sec:background}

\subsection{Sharding in Databases}
Traditional database systems achieve scalability by dividing the database states
into independent shards (or partitions).
By distributing the workload over multiple shards, the overall capacity of the
system increases.
Sharding requires coordination to ensure ACID properties for transactions that access
multiple shards.  
Two important coordination protocols are distributed commit such as two-phase commit (2PC) which ensures atomicity,
and concurrency control such as two-phase locking (2PL) which achieves isolation. 

In this paper, we use the term {\em sharding} to refer to the combination of
replication and partitioning as shown in \autoref{fig:shard_db}.  This
architecture is adopted by recent distributed database systems
to achieve fault tolerance and 
scalability~\cite{google_spanner}. Each partition is replicated over multiple replicas, and its
content is kept consistent by consensus protocols~\cite{paxos, raft}. 
Transaction management and consensus protocols can be combined
more judiciously, as opposed to layering one on top of another, to achieve
better performance~\cite{tapir,janus,carousel}.

Sharding in database systems assumes crash-failure model, in which a faulty node
stops sending and responding to requests. There are three important implications
of this assumption.  First, efficient consensus protocols catered for
crash-failure model can be used to achieve high performance.
Second, creating a shard is simple.  For example, a node can be assigned to a
shard based on its location.  Third, the coordinators that drive coordination
protocols are fully trusted.

\subsection{Blockchains Consensus Protocols}
\label{subsec:background_consensus}
Blockchain is essentially a distributed, append-only ledger that stores a
sequence of transactions.
The blocks are chained together by cryptographic hash pointers.
The blockchain is maintained by a set of mutually distrusting nodes (a.k.a.
replicas or validators). These nodes run a consensus protocol to ensure the
blockchain's consistency under Byzantine (or arbitrary) failures. This is in
contrast to distributed databases whose threat model does not account for
Byzantine failures or malicious users~\cite{untangling_blockchain}.

Blockchain consensus protocols should achieve both {\em safety} and {\em liveness} despite Byzantine failures. 
Safety means that honest (non-Byzantine) nodes agree on the same value, whereas liveness means
that these nodes eventually agree on a value.  Two major classes of blockchain consensus protocols are Byzantine Fault
Tolerance and Nakamoto consensus.

{\bf Byzantine Fault Tolerant (BFT) protocols.} Practical Byzantine Fault
Tolerance (PBFT)~\cite{pbft}, the most well-known BFT protocol, consists of
three phases: a {\em pre-prepare} phase in which the leader broadcasts requests
as \texttt{pre-prepare} messages, the {\em prepare} phase in which replicas
agree on the ordering of the requests via \texttt{prepare} messages, and the
{\em commit phase} in which replicas commit to the requests and their order via
\texttt{commit} messages.
Each node collects a quorum of \texttt{prepare} messages before moving to the
commit phase, and executes the requests after receiving a quorum of
\texttt{commit} messages. A faulty leader is replaced via the {\em view change}
protocol.
The protocol uses $O(N^2)$ messages for $N$ replicas. For $N \geq 3f+1$, it
requires a quorum size of $2f+1$ to tolerate $f$ failures. It achieves safety in
asynchronous networks, and liveness in partially synchronous networks wherein
messages are delivered within an unknown but finite bound. More recent BFT
protocols~\cite{xft} extend PBFT to optimize for the normal case (without view
change).


{\bf Nakamoto consensus protocols.} Proof-of-Work (PoW)~\cite{btc_origin}, as
used in Bitcoin, is the most well-known instance of Nakamoto consensus.
The protocol randomly selects a leader to propose the next block.
Leader selection is a probabilistic process in which a node must solve a
computational puzzle to claim leadership.  The probability of solving the puzzle
is proportional to the amount of computational power the node possesses over the
total power of the network.  The protocol suffers from forks which arise when
multiple nodes proposes blocks roughly at the same time. It has low throughput,
but can scale to a large number of nodes.

Nakamoto consensus protocols quantify Byzantine tolerance in terms of the cumulative resources belonging to
the Byzantine nodes (e.g., fraction of the total computational power). Their safety depends not only on the
Byzantine threshold, but also on network latency.  Under a fully synchronous network, safety is guaranteed
against $50\%$ Byzantine attackers~\cite{btc_origin}.  However, this threshold drops quickly in a partially
synchronous network, going below $33\%$ when the latency is equal to the block time~\cite{bitcoin_async}.

\subsection{Trusted Execution Environment (TEE)}
\label{subsec:sgx}
One approach to improve BFT protocols  is to assume a 
hybrid failure model in which some components are trusted and only fail by
crashing, while others behave in a Byzantine manner.  This model is realized by
running the trusted components inside a TEE.  One important security guarantees
of TEE is that it ensures the integrity of the protected components, so that the
attackers cannot tamper with their execution and cause them to deviate from the prescribed protocols.
Our work uses Intel Software Guard Extensions (SGX)~\cite{sgx} to provision the TEE. But we note that our
design can work with other TEE instantiations (e.g., TrustZone~\cite{trustzone}, Sanctum~\cite{sanctum}).

{\bf Enclave protection.} Intel SGX provides TEE support in the form of {\em
hardware enclave}.
An enclave is a CPU-protected address space which is accessible \textit{only} by
the code within the enclave (i.e., the trusted component).
Multiple enclaves can be instantiated by non-privileged user processes. The
enclaves are isolated from each other, from the operating system (OS) and from
other user processes. The enclave code, however, can invoke OS services such as
IO and memory management.

{\bf Attestation.} A user can verify if a specific TEE is correctly instantiated
and running at a remote host via a remote attestation
protocol~\cite{sgx_remote_attest}. Once the enclave in question has been
initiated, the CPU computes a \textit{measurement} of this enclave represented
by the hash of its initial state. The CPU then signs the measurement with its
private key. The user can verify the signed message, and then compare the
measurement against a known value.

{\bf Data sealing.} The TEE can persist its state to non-volatile memory via the data sealing
mechanism, which allows for recovery after crash failures. An enclave {\em
seals} its data by first requesting the CPU for a unique key bound to its measurement, then encrypting the data before
storing it on persistent storage. This mechanism ensures the data can only be decrypted by the enclave that
sealed it. However, enclave recovery is subject to rollback attacks wherein an attacker (e.g., the malicious OS) provides properly sealed but stale data to the enclave~\cite{matetic2017rote}.

{\bf Other cryptographic primitives.} {SGX provides $\texttt{sgx\_read\_rand}$ and
$\texttt{sgx\_get\_trusted\_time}$ functions to the enclave processes. The former generates unbiased random numbers, the latter returns the elapsed time relative to a reference point.}

\section{Overview}
\label{sec:problem}
In this section, we discuss our goals and illustrate them with a running
example. We detail the challenges in realizing these goals and present our
sharding approach. Finally, we describe the system model and security
assumptions.


\subsection{Goals}
The design of a highly scalable blockchain must meet the following three goals:
(i) support a large network size equivalent to that of major cryptocurrencies
like Bitcoin and Ethereum, (ii) achieve high transaction throughput that can
handle the average workloads of centralized systems such as Visa, and (iii)
support general workloads and applications beyond cryptocurrencies.  The
resulting blockchain will enable scale-out applications that offer the security
benefits of a decentralized blockchain with a performance similar to that of a
centralized system. To better motivate and illustrate our design, we use the
following example throughout the paper.

\noindent{\bf Running example.} Consider a consortium of financial institutions that offer cross-border financial
services to their customers. They implement a blockchain solution that provides a shared ledger for recording
transactions which can be payments, asset transfers or settlements. Unlike Bitcoin or Ripple~\cite{ripple},
there is no native currency involved. The ledger is distributed and its content is agreed upon by consortium
members via distributed consensus.  Given the amount of money at stake, the blockchain solution must be
tolerant to Byzantine failures, so that group members are protected against attacks that compromise the ledger
in order to double-spend or to revoke transactions. As the consortium can admit new members, the system should
not assume any upper bound on the consortium size. Finally, the blockchain must be able to support high
request rates and deliver high transaction throughput.

\subsection{Challenges and Approach}
Building a blockchain system that achieves all three goals above at the same
time is challenging. To have high transaction throughput (second goal), it is
necessary to build on top of a permissioned blockchain. But such a blockchain
uses BFT protocols which do not scale to a large number of nodes, thus
contradicting the first goal. As a result, one challenge is to reconcile the
first two goals by making BFT protocols more scalable. We note that scalability
here means fault scalability, which means that protocol's performance degrades
gracefully as the number of tolerated failures increases. We address this
challenge by using trusted hardware to remove the attacker's ability to
equivocate. Specifically, we use a hardware-assisted PBFT protocol that requires
only $N = 2f + 1$ replicas to tolerate $f$ failures. We implement this protocol
on top of Hyperledger's PBFT implementation and further improve its scalability
by introducing two protocol-level optimizations that reduce the number of
message broadcasts, and an implementation-specific optimization that avoids
dropping important messages.

We cannot achieve the first two goals by improving BFT protocols alone, because
there is a limit on scalability due to the quadratic communication cost of
$O(N^2)$. Our approach is to apply the database sharding technique to partition
the network and the blockchain states into smaller shards, where each shard is
small enough to run a BFT protocol. In distributed databases, nodes can be
assigned to shards randomly or based on their location. But in blockchain, the
process of assigning nodes to shards, called {\em shard formation}, must be done
carefully to ensure security because each shard can only tolerate a certain
number of Byzantine failures. In particular, given a network of $N$ nodes, a
fraction $s$ of which are Byzantine, shard formation must guarantee with
overwhelming probability that no shard of size $n \ll N$ contains more than $f$
Byzantine nodes over the entire lifetime of the system\footnote{It must be the
case that $\frac{f}{n} \leq s$ in order for this requirement to be met.}. The
relationship between $f$ and $n$ depends on the consensus protocol. The
challenge here is to perform shard formation securely and efficiently. Existing
solutions use expensive cryptographic protocols and the resulting shards are
large. In contrast, we leverage TEE to implement an efficient trusted randomness
beacon that serves as the random source for the shard formation.  Furthermore,
our fault-scalable PBFT protocol allows for shards of smaller size.  This leads
to higher throughput per shard, and more shards given the same network size.

As in distributed databases, sharding requires coordination protocols for
cross-shard (or distributed) transactions.
Two important properties of coordination protocols are {\em safety} which means
atomicity and isolation, and {\em liveness} which means the transaction will
eventually abort or commit.  The challenge in realizing our third goal is to
design a coordination protocol that supports non-UTXO distributed transactions,
while achieving safety and liveness even when the coordinator is malicious.
Existing sharded blockchains~\cite{elastico, omniledger, rapidchain} do not fully address this challenge, as we elaborate later in
Section~\ref{sec:coordinating_shard}.  Our
approach is to use the classic 2PC and 2PL protocols to ensure safety, and run 2PC in a Byzantine tolerant
shard to avoid malicious blocking.  This coordination protocol works for all
blockchain data models and applications.

The key components of our design are summarized in
\autoref{fig:shard_bc}.  First, our shard formation protocol securely partitions
the network into multiple committees, thereby allowing the system throughput to
scale with the number of nodes in the system. This protocol relies on a trusted
randomness beacon implemented inside a TEE for efficiency.  Second, each shard
runs our scalable BFT protocol which achieves high throughput at scale by
combining TEE with other optimizations. Finally, layered on top of the shards is
a distributed transaction protocol that achieves safety and liveness for general
blockchain applications.

\subsection{System and Threat Model}
\label{subsec:threat_model}
\textbf{System model.} We consider a blockchain system of $N$ nodes, with a
fraction $s$ of the network under the attacker's control, while the remaining
fraction is honest.
The shard formation protocol partitions the nodes into $k$ committees, each
consisting of $n \ll N$ nodes. Each committee can tolerate at most $f < n$
Byzantine nodes.
The committees maintain disjoint partitions of the blockchain states (i.e.,
shards).
Unless otherwise stated, the network is partially synchronous,
in which messages sent repeatedly with a finite time-out will eventually be
received. This is a standard assumption in existing blockchain
systems~\cite{omniledger, elastico}.

In the running example above, suppose the consortium comprises $400$
institutions, among which $100$ members actively collude so that they can revoke
transactions that transfer their assets to the remaining institutions. In such
a case, $N=400$ and $s=25\%$. Suppose further that the consortium partitions
their members into four equally-sized committees, then $n=100$. Each committee
{\em owns} a partition of the ledger states.  The committee members run a
consensus protocol to process transactions that access the committee's states.
If PBFT is used, each committee can tolerate at most $f=\frac{n-1}{3}=33$ Byzantine nodes.

Every node in the system is provisioned with TEEs. We leverage
Intel SGX in our implementations, but our design can work with any other TEE
instantiations, for example hardware-based TEEs such as
TrustZone~\cite{trustzone}, Sanctum~\cite{sanctum}, TPMs~\cite{TPM}, or
software-based TEEs such as Overshadow~\cite{overshadow}.

\noindent{\bf Threat model.} 
The attacker has full control of the Byzantine nodes. It can read and write to the memory of
any running process, even the OS. It can modify data on disk, intercept and change the content of any system
call. It can modify, reorder and delay network messages arbitrarily. It can start, stop and invoke the local
TEE enclaves with arbitrary input. However, its control of the enclaves is restricted by the TEE threat model
described below. The attacker is adaptive, like in Elastico~\cite{elastico} and OmniLedger~\cite{omniledger},
meaning that it can decide which honest nodes to corrupt. However, the corruption does not happen instantly,
like in Algorand~\cite{algorand}, but takes some time to come into effect. Furthermore, the attacker can only
corrupt up to a fraction of $s$ nodes at a time. It is computationally bounded and cannot break standard
cryptographic assumptions. Finally, it does not mount denial-of-service attacks against the system.



The threat model for TEE is stronger than what SGX currently
offers. In particular, SGX assumes that the adversary cannot compromise the
integrity {\em and} confidentiality of protected enclaves.
For TEE, we also assume that integrity protection mechanism is secure. But there
is no guarantee about confidentiality protection, except for a number of
important cryptographic primitives: attestation, key generation, random number
generation, and signature. In other words, enclaves have no private memory
except for areas related to its private keys, i.e., they run in a seal-glassed
proof model where their execution is transparent~\cite{sealed_glass_proof}. This
model admits recent side-channel attacks on SGX that leak enclave
data~\cite{grandexp}. Although attacks that leak attestation and other private
keys are excluded~\cite{foreshadow}, we note that there exist both software and
hardware techniques to harden important cryptographic operations against side
channel attacks.

\section{Scaling Consensus Protocols}
\label{sec:existing_protocols}

\subsection{Scaling BFT Consensus Protocol}
\label{subsec:ahl}
PBFT, the most prominent instance of BFT consensus protocols, has been shown not
to scale beyond a small number of nodes due to its communication
overhead~\cite{blockbench}. In the running example, this means each committee in
the consortium can only comprise dozens of institutions.  Furthermore, the
probability of the adversary controlling more than a third of the committee is
high when the committee size is small. Our goal is to improve both the
protocol's communication overhead and its fault tolerance.

\begin{figure}
\centering
\includegraphics[width=0.48\textwidth]{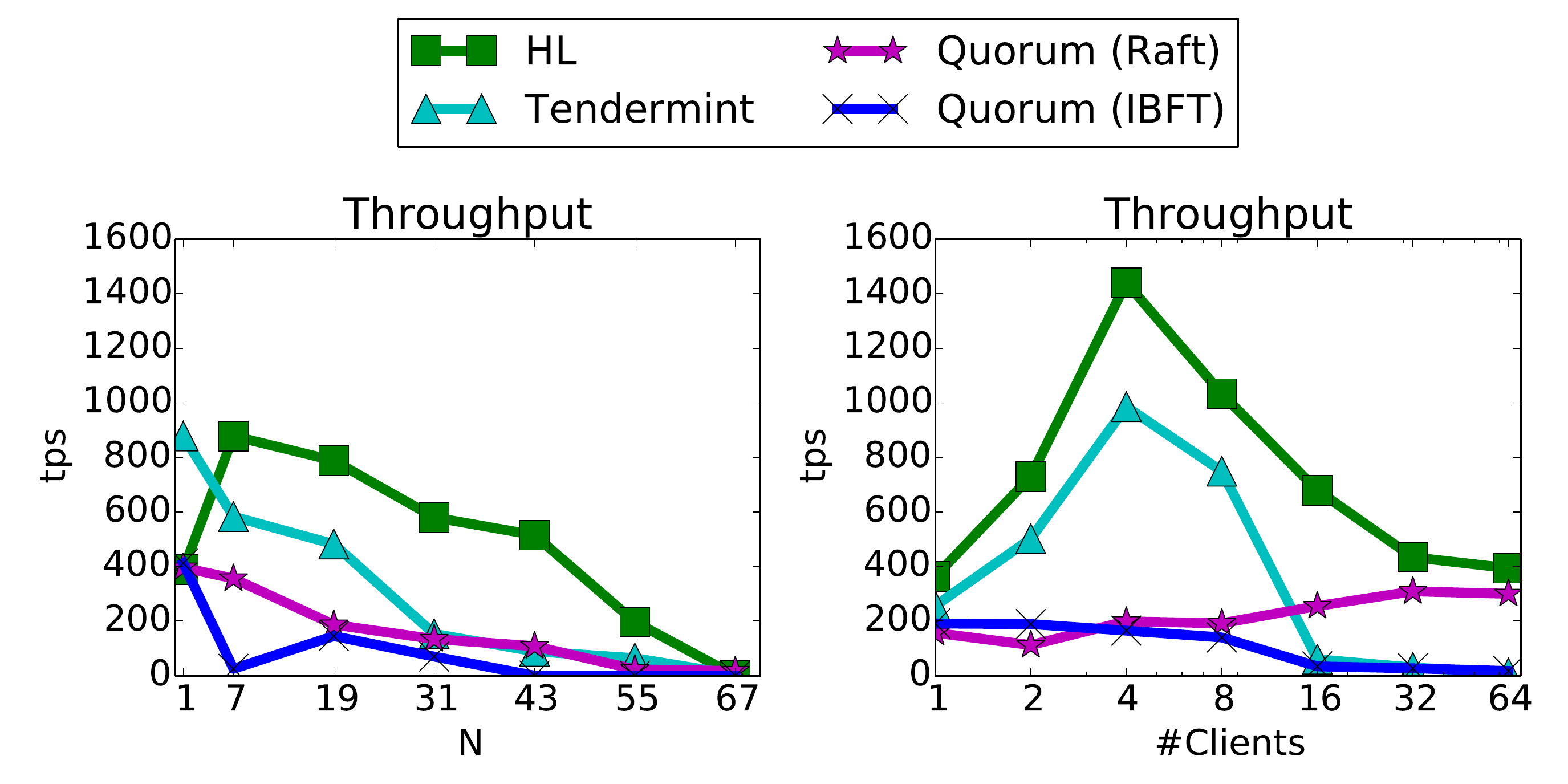}
\caption{Comparison of BFT protocols with varying number
of nodes and clients.}
\label{fig:cluster_bft}
\end{figure}

\vspace{2mm}
\noindent\textbf{Why PBFT?} There are several BFT
implementations for blockchains. PBFT is adopted by Hyperledger.
Tendermint~\cite{tendermint} -- a variant of PBFT -- is used by Ethermint and
Cosmo. Istanbul BFT (IBFT) is adopted by Quorum. Raft~\cite{raft}, which only
tolerates crash failures, is implemented by Coco to tolerate Byzantine failures
by running the entire protocol inside Intel SGX~\cite{mscoco}.
Figure~\ref{fig:cluster_bft} compares the throughputs of these BFT
implementations, where we use Raft implementation in Quorum as an approximation
for Coco whose source code is not available.  Due to space constraint, we only
highlight important results here, and include detailed discussion in
Appendix~\ref{appendix_subsec:bfts}.  PBFT consistently outperforms the
alternatives at scale. The reason is that PBFT design permits pipelined
execution, whereas IBFT and Tendermint proceed in lockstep. Although pipelined
execution is possible in Raft, this property is not exploited in Quorum. From
this observation, we base our sharded blockchain design on top of Hyperledger,
and focus on improving PBFT.

\noindent\textbf{Reducing the number of nodes.} If a consensus protocol can prevent Byzantine nodes
from equivocating (i.e., issuing conflicting statements to different nodes), it
is possible to tolerate up to $f=\frac{N-1}{2}$ non-equivocating Byzantine failures out of $N$ nodes~\cite{a2m}.
Equivocation can be eliminated by running the entire consensus protocol
inside a TEE, thereby reducing the failure model from Byzantine to crash failure~\cite{mscoco}. 
We do not follow this approach, however, because it incurs a large trusted code base (TCB). A large TCB is
undesirable for security because it is difficult, if not impossible to conduct security analysis for the code
base, and it increases the number of potential vulnerabilities~\cite{sgx_guide}. 
Instead, we adopt the protocol proposed by Chun {\em et al.}~\cite{a2m} which uses a small trusted log
abstraction called attested append-only memory to remove equivocation.  The log is maintained inside the TEE
so that the attacker cannot tamper with its operations.  We implement this protocol on top of Hyperledger
Fabric v0.6 using SGX, and call it AHL (\underline{A}ttested \underline{H}yper\underline{L}edger).

AHL maintains different logs for different consensus message types (e.g.,
\texttt{pre-prepare}, \texttt{prepare}, \texttt{commit}). Before sending out a
new message, each node has to append the message's digest to the corresponding
log. The proof of this operation contains the signature created by the TEE, and
is included in the message.  AHL requires all valid messages to be accompanied
by such  proof.  Each node collects and verifies $f+1$ \texttt{prepare} messages
before moving to the commit phase, and $f+1$ \texttt{commit} messages before
executing the request. AHL periodically seals the logs and writes them to
persistent storage. This mechanism, however, does not offer protection against
rollback attacks~\cite{matetic2017rote}.
We describe how to extend AHL to guard against these attacks in
Appendix~\ref{apd:rollback_attacks}.


\vspace{2mm}
\noindent\textbf{Optimizing communications.} 
Our evaluation of AHL in 
Section~\ref{sec:eval} shows that it fails to achieve the desired scalability.  We observe a high number of
consensus messages being dropped, which leads to low throughput when the network size increases.  From this
observation, we introduce two optimizations to improve the communications of the system, and refer to the
resulting implementation as AHL+.

First, we improve Hyperledger implementation which uses the same network queue for both consensus and
request messages.  In particular, we split the original message queue into two separated
channels, each for a different type of message. Messages received from the network contain metadata that
determines their types and are forwarded to the corresponding channel. This separation prevents request
messages from overwhelming the queue and causing consensus messages to be dropped.

Second, we note that when a replica receives a user request, the PBFT protocol specification states that the
request is broadcast to all nodes~\cite{pbft_thesis}.  However, this is not necessary as long as the request
is received at the leader, for the leader will broadcast the request again during the {\em pre-prepare} phase.
Therefore, we	 remove the request broadcast. The replica receiving the request
from the client simply forwards it to the leader. We stress that this is a design-level optimization.


We also consider another optimization adopted by Byzcoin~\cite{byzcoin}, in which the leader 
collects and aggregates other nodes' messages into a single authenticated message. Each node forwards its
messages to the leader and verifies the aggregated message from the latter. As a result, the communication
overhead is reduced to $O(N)$. This design, called AHLR (\underline{A}ttested \underline{H}yper\underline{L}edger
\underline{R}elay), is implemented on top of AHL via an enclave that verifies and
aggregates messages. Given valid $f+1$ signed messages for a request $\textit{req}$, in phase $p$ of consensus
round $o$, the enclave issues a proof indicating that there has
been a quorum for $\langle \textit{req},p,o \rangle$. 

\vspace{2mm}
\noindent \textbf{Security analysis.} The trusted log
operations in AHL are secure because they are signed by private keys generated
inside the enclave. Because the adversary cannot forge signatures of the logs'
operations, it is not possible for the Byzantine nodes to equivocate. As shown
in~\cite{a2m}, given no more than $f=\frac{N-1}{2}$ non-equivocating Byzantine
failures, AHL guarantees safety regardless of the network condition, and
liveness under partially synchronous network.
AHL+ only optimizes communication between nodes and does not change the
consensus messages, therefore it preserves AHL properties. AHLR only optimizes
communication in the normal case when there is no view change, and uses the same
view change protocol as in AHL. Because message aggregation is done securely
within the enclave, AHLR has the same safety and liveness guarantees as AHL.


\subsection{Scaling \poet\ Consensus Protocol}
\label{subsec:poet}
Proof of Elapsed Time (\poet) is a variant of Nakamoto consensus, wherein nodes
are provisioned with SGXs.
Each node asks the enclave for a randomized {\em \texttt{waitTime}}. 
Only after such \texttt{waitTime} expires does the enclave issue a {\em wait certificate} or create a new
\texttt{waitTime}.  The node with the shortest \texttt{waitTime} 
becomes the leader and is able to propose the next block.

Similar to PoW, \poet~suffers from forks and stale blocks.  
Due to propagation delays, if multiple nodes obtain their certificates 
roughly at the same time, they will propose conflicting blocks, 
creating forks in the blockchain. 
The fork is resolved based on the aggregate resource contributed to the branches,
with blocks on the losing branches discarded as stale blocks. 
Stale block rate has a negative impact on both the security and throughput of the
system~\cite{pow_sec_vs_tps}.


\vspace{2mm}
\noindent\textbf{\spoet: Improving PoET.} 
We improve \poet\ by restricting the number of nodes competing to propose the
next block,  thereby reducing the stale block rate.  We call this optimized protocol \spoet.  Unlike \poet,
when invoked to generate a wait certificate, \spoet\ first uses {\tt sgx\_read\_rand} to generate a random
$l$-bit value \texttt{q} that is bound to the wait certificate.  Only wait certificates with $\texttt{q}=0$
are considered valid.  The node with a valid certificate and the shortest \texttt{waitTime} becomes the
leader.  \spoet\ leader selection can thus be seen as a two-stage
process. The first stage samples uniformly at random a subset of $n' = n \cdot
2^{-l}$ nodes. The second stage selects uniformly at random a leader among these
$n'$ nodes. It can be shown that the expected number of stale blocks in \spoet\ is smaller than that in \poet.

\vspace{2mm}
\noindent\textbf{\spoet\ vs AHL+.}
\spoet\ safety depends not only on the Byzantine threshold, but also on network latency.  In a partially
synchronous network, its fault tolerance may drop below $33\%$~\cite{pow_sec_vs_tps}.  This is in contrast to
AHL+ whose safety does not depend on network assumption.  More importantly, our performance evaluation of
\spoet\ (included in
Appendix ~\ref{sec:apd_experiments}) shows that it has lower throughput than AHL+. Therefore, we adopt AHL+
for the design and implementation of the sharded blockchain.  

\vspace{2mm}

\section{Shard Formation}
\label{sec:shard_formation}
{ Forming shards in a blockchain system is more complex than in
a distributed database. First, the nodes must be assigned to committees in an
unbiased and random manner. Second, the size of each committee must be selected
carefully to strike a good trade-off between performance and security. And
finally, committee assignment must be performed periodically to prevent an
adaptive attacker from compromising a majority of nodes in a committee. This
section presents our approach of exploiting TEEs to address these challenges.
}
\subsection{Distributed Randomness Generation} 
\label{subsec:rand_gen}
{ A secure shard formation requires an unbiased random number 
{\tt rnd} to seed the node-to-committee assignment. Given {\tt rnd}, the nodes
derive their committee assignment by computing a random permutation $\pi$ of
$[1:N]$ seeded by {\tt rnd}.  $\pi$ is then divided into approximately
equally-sized chunks, each of which represents the members in one committee.}


We exploit TEEs to efficiently obtain {\tt rnd} in a distributed and Byzantine
environment, by equipping each node with a \textproc{RandomnessBeacon} enclave
that returns fresh, unbiased random numbers.
Similar to prior works~\cite{omniledger, elastico, rapidchain}, we assume a
synchronous network with the bounded delay $\Delta$ during the distributed
randomness generation procedure.

Our sharded blockchain system works in epochs. 
Each new epoch corresponds to a new node-to-committee assignment. 
At the beginning of each
epoch, each node invokes the \textproc{RandomnessBeacon} enclave with an epoch number $e$. The enclave generates two random
values {\tt q} and {\tt rnd} using two independent invocations of the {\tt
sgx\_read\_rand} function.  It then returns a signed certificate containing
$\langle e, \texttt{rnd}\rangle$ if and only if $\texttt{q} = 0$.  The
certificate is broadcast to the network. After a time $\Delta$,  nodes lock
in the lowest {\tt rnd} they receive for epoch $e$, and uses it to compute the
committee assignment.


The enclave is configured such that it can only be invoked once per epoch, which is to prevent
the  attacker from selectively discarding the enclave's output in order to bias the final randomness. If the
nodes fail to receive any message after $\Delta$, which happens when no node can obtain  $\langle e,
\texttt{rnd}\rangle$ from its enclave, they increment $e$ and repeat the process. The probability of repeating
the process is $P_{\text{repeat}} = (1-2^{-l})^N$ where $l$ is the bit length of {\tt q}.  It can be tuned to
achieve a desirable trade-off between $P_{\text{repeat}}$ and the communication overhead, which is $O(2^{-l}
N^2)$.  For example, when $l=\log(z)$ for some constant $z$, $P_{\text{repeat}} \approx 0$ and the
communication is $O(N^2)$. When $l=\log(N)$, $P_{\text{repeat}} \approx {\mathrm{e}}^{-1}$ and the communication is
$O(N)$.

\vspace{0.2cm}
\noindent \textbf{Security analysis.} Because $\texttt{q}$ and $\texttt{rnd}$ are generated independently
inside a TEE, their randomness is not influenced by the attacker. Furthermore, the enclave only
generates them once per epoch, therefore the attacker cannot selectively discard the outputs to bias the
final result and influence the committee assignment.

\subsection{Committee Size}
\label{subsec:comm_size}
{Since committee assignment is determined by a random
permutation $\pi$ of $[1:N]$ seeded by {\tt rnd}, it can be seen as random
sampling without replacement.  Therefore, we can compute the probability of a
faulty committee (i.e., a committee containing more than $f$ Byzantine nodes)
using the hypergeometric distribution.  In particular, let $X$ be a random
variable that represents the number of Byzantine nodes assigned to a committee
of size $n$, given the overall network size of $N$ nodes among which up to
$F=sN$ nodes are Byzantine. The probability of faulty committee, i.e., the
probability that security is broken, is:
}
\begin{equation} Pr [X \geq f] = \sum_{x=f}^{n}{\frac{{{F}\choose{x}} {{N-F}\choose{n-x}} }{{{N}\choose{n}}}}
\label{eq:comm_fault}
\end{equation}

%

\noindent\textbf{Keeping the probability of faulty committee negligible.}
{We can bound the probability of faulty committee to be
negligible by carefully configuring the committee size, based on
Equation~\ref{eq:comm_fault}.
If $f \leq \frac{n-1}{3}$ (as in the case of PBFT), in the presence of a $25\%$
adversarial power, each committee must contain $600+$ nodes to keep the faulty
committee probability negligible (i.e., $Pr [X \geq \frac{n-1}{3}] \leq
2^{-20}$).  When AHL+ is used, each committee can tolerate up to $f =
\frac{n-1}{2}$, thus the committees can be significantly smaller: $n = 80$ for
$Pr [X \geq \frac{n-1}{2}] \leq 2^{-20}$.
}

Smaller committee size leads to better performance for two reasons. First, individual committees achieve
higher throughput due to lower communication overhead.  Second, there are more committees in the network,
which can increase throughput under light contention workloads. We report the committee sizes with respect to
different adversarial power and their impact on the overall throughput in Section~\ref{sec:eval}.

\subsection{Shard Reconfiguration}
\label{subsec:shard_reconfig}
{ An adaptive attacker may compromise a non-faulty committee by
corrupting otherwise honest nodes. Our threat model, however, assumes that such
node corruption takes time. As a result, we argue that periodic committee
re-assignment, or shard reconfiguration, that reshuffles nodes among committees,
suffices to guard the system against an adaptive attacker.}

{ Shard reconfiguration occurs at every epoch. At the end of
epoch $e-1$, nodes obtain the random seed {\tt rnd} following the protocol
described in Section~\ref{subsec:rand_gen}. They compute the new committee
assignment for epoch $e$ based on {\tt rnd}.  We refer to nodes whose committee
assignment changes as {\it transitioning nodes}. We refer to the period during
which transitioning nodes move to new committees as the \textit{epoch
transition} period.
}

{ During epoch transition, transitioning nodes first stop
processing requests of their old committees, then start fetching the states of
their new committees from current members of the corresponding committees. Only
after the state fetching completes do they officially join the new committee and
start processing transactions thereof. During this period, the transitioning
nodes do not participate in the consensus protocol of either their old or new
committees.  Consequently, a naive reconfiguration approach in which all nodes
transition at the same time is undesirable, as it renders the system
non-operational during the transition period.}

{Our approach is to have nodes transitioning in batches. In
particular, for each committee, only up to {\tt B} nodes move to new committees
at a time. The order by which nodes move is determined based on {\tt rnd}, which
is random and unbiased.  In the following, we reason about the impact of {\tt B}
on the safety and liveness of the sharded blockchain.  }

\vspace{2mm}
\noindent\textbf{Safety analysis.} { Let $k$ be the number of
shards, where each shard represents a partition of the global blockchain states.
A shard reconfiguration essentially changes the set of nodes that processes
requests for each of the $k$ shards.
Consider a shard {\tt sh}, and denote the committee handling {\tt sh} in epoch
$e-1$ by $C_{e-1}$ and in epoch $e$ by $C_e$. Since {\tt B} nodes are switching
out of ${\texttt{C}}_{e-1}$ at a time, and there are $\frac{n}{k}$ nodes of
${{C}}_{e-1}$ expected to remain in $C_e$, there are
$\frac{n(k-1)}{k\cdot\texttt{B}}$ {\em intermediate} committees handling {\tt
sh} during the epoch transition period.}

{ Swapping out {\tt B} nodes does not violate safety of {\tt sh}, because the number of
Byzantine nodes in the current committee does not increase. On the other hand, when new  {\tt B} nodes are
swapped in, the number of Byzantine nodes in the intermediate committee may exceed the safety threshold. As
the transitioning nodes are chosen at random based on {\tt rnd}, the probability of the intermediate
committee being faulty follows Equation~\ref{eq:comm_fault}. In expectation, there are
$\frac{n(k-1)}{k\cdot\texttt{B}}$ such intermediate committees during the transitioning from
${{C}}_{e-1}$ to $C_e$. We use Boole's inequality to estimate the probability
that the safety of shard {\tt sh} is violated during the epoch
transitioning:} \begin{equation} Pr (\texttt{faulty}) \leq
\sum_{i=1}^{\frac{n(k-1)}{k\cdot\texttt{B}}}{\sum_{x=f}^{n}{\frac{{{F}\choose{x}} {{N-F}\choose{n-x}}
}{{{N}\choose{n}}}}} \label{eq:epoch_transition_fault}
\end{equation}
{
For example, with $n=80$, $f=\frac{n-1}{2}$, $k=10$ shards, and $\texttt{B}=\log(n) = 6$, $Pr (\texttt{faulty}) \approx 10^{-5}$.
Based on Equation~\ref{eq:epoch_transition_fault}, we can configure $\texttt{B}$ to balance between liveness and safety of the system during epoch transition.
}

\vspace{2mm}
\noindent\textbf{Liveness analysis.} {During the transitioning,
each committee has {\tt B} nodes not processing requests. If $\texttt{B} > f$,
the shard cannot make progress because the remaining nodes cannot form a quorum.
Thus, the larger {\tt B} is, the higher the risk of loss of liveness during
epoch transition. }

\section{Distributed Transactions}
\label{sec:coordinating_shard}


In this section, we explain the challenges in supporting
distributed, general transactions for blockchains. We discuss the limitations of
state-of-the-art systems:
RapidChain~\cite{rapidchain} and OmniLedger~\cite{omniledger}
(Elastico~\cite{elastico} is not considered because it does not support
distributed transactions). We then present a solution that enables
fault-tolerant, distributed, general transactions, and discuss how it can be
improved.  

\subsection{Challenges}
In a sharded blockchain, a distributed (or cross-shard)
transaction is executed at multiple shards.
Appendix~\ref{apd:cross_shard_prob} shows that in practical blockchain
applications, a vast majority of transactions are distributed. Similar to
databases, supporting distributed transactions is challenging due to the safety
and liveness requirements.  The former means atomicity and isolation that handle
failures and concurrency, the latter means that transactions do not block
forever. We note that in the sharded blockchain, concurrency does not arise
within a single shard, because the blockchain executes transaction sequentially.
Instead, as we explain later, concurrency arises due to cross-shard
transactions.

{\bf UTXO transactions.} Bitcoin and many other
cryptocurrencies adopt the Unspent Transaction Output (UTXO) data model. A UTXO
transaction consists of a list of inputs, and a list of outputs. All the inputs
must be the outputs of previous transactions that are unspent (i.e., they have
not been used in another transaction). The outputs of the transaction are new,
unspent coins. Given a transaction, the blockchain nodes check that its inputs
are unspent, and the sum of the outputs is not greater than that of the inputs.
If two transactions consume the same unspent coins, only one is accepted.

\def\hd{2}
\def\vd{3}
\begin{figure}
\centering
\begin{subfigure}{.49\textwidth}
\centering    
\resizebox{.96\textwidth}{!}{  
\begin{tikzpicture}[scale=0.6, every text node part/.style={align=center}]  
\node[circle, draw = amaranth, fill = amaranth, scale=0.6] (C1){C};
\node[rectangle, draw = airforceblue, fill = airforceblue, below of = C1, xshift = -1*\hd em, node distance = \vd em] (S11){$S_1$};
\node[rectangle, draw = asparagus, fill = asparagus, below of = C1, node distance = \vd em] (S12){$S_2$};
\node[rectangle, draw = bronze, fill = bronze, below of = C1, xshift = \hd em, node distance = \vd em] (S13){$S_3$};
\draw[->] (C1.south east) -- node[xshift = 0.5em, scale = 0.8, yshift = 0.25em] {$tx$}(S13.north);
\node[circle, draw = amaranth, fill = amaranth, scale=0.6, right of = C1, node distance = 7*\hd em] (C2){C};
\node[rectangle, draw = airforceblue, fill = airforceblue, below of = C2, xshift = -1*\hd em, node distance = \vd em] (S21){$S_1$};
\node[rectangle, draw = asparagus, fill = asparagus, below of = C2, node distance = \vd em] (S22){$S_2$};
\node[rectangle, draw = bronze, fill = bronze, below of = C2, xshift = \hd em, node distance = \vd em] (S23){$S_3$};

\draw[->] (S23.north) -- ($(S23.north)+(0,0.5)$) -| node[yshift = 0.5em, scale = 0.8] {$tx_a$}(S21.north);
\draw[->] (S23.south) -- ($(S23.south)+(0,-0.5)$) -| node[xshift = -1em, scale = 0.8] {$tx_b$}(S22.south);
\node[circle, draw = amaranth, fill = amaranth, scale=0.6, right of = C2, node distance = 7*\hd em] (C3){C};
\node[rectangle, draw = airforceblue, fill = airforceblue, below of = C3, xshift =  -1*\hd em, node distance = \vd em] (S31){$S_1$};
\node[rectangle, draw = asparagus, fill = asparagus, below of = C3, node distance = \vd em] (S32){$S_2$};
\node[rectangle, draw = bronze, fill = bronze, below of = C3, xshift = \hd em, node distance = \vd em] (S33){$S_3$};

\draw[->] (S31.north) -- node[yshift = 1em, scale = 0.8] {$I'_1$} ($(S31.north)+(0,0.5)$) -| ($(S33.north)+(-0.25,0)$);
\draw[->] (S32.south) -- ($(S32.south)+(0,-0.5)$) -| node[xshift = -2.5em, scale = 0.8] {$I'_2$}(S33.south);
\draw[->, densely dashed] ($(S33.north)+(0.25,0)$) |-node[xshift = 0em, yshift = -1em, scale = 0.8] {$I'_1$, if \\ \ \ \ $I'_2=\bot$} (C3.east);
\end{tikzpicture}
}
\caption{RapidChain}
\label{subfig:Rapidchain_cross_shard}
\end{subfigure}
\begin{subfigure}{.49\textwidth}
\centering    
\resizebox{.96\textwidth}{!}{  
\begin{tikzpicture}[scale=0.6, every text node part/.style={align=center}]  
\node[circle, draw = amaranth, fill = amaranth, scale=0.6] (C1){C};
\node[rectangle, draw = airforceblue, fill = airforceblue, below of = C1, xshift = -1*\hd em, node distance = \vd em] (S11){$S_1$};
\node[rectangle, draw = asparagus, fill = asparagus, below of = C1, node distance = \vd em] (S12){$S_2$};
\node[rectangle, draw = bronze, fill = bronze, below of = C1, xshift = \hd em, node distance = \vd em] (S13){$S_3$};
\draw[->] (C1.south west) -- node[xshift = -0.5em, scale = 0.8] {$tx$}(S11.north);
\draw[->] (C1.south) -- node[xshift = 0.5em, scale = 0.8] {$tx$}(S12.north);
\node[circle, draw = amaranth, fill = amaranth, scale=0.6, right of = C1, node distance = 7*\hd em] (C2){C};
\node[rectangle, draw = airforceblue, fill = airforceblue, below of = C2, xshift = -1*\hd em, node distance = \vd em] (S21){$S_1$};
\node[rectangle, draw = asparagus, fill = asparagus, below of = C2, node distance = \vd em] (S22){$S_2$};
\node[rectangle, draw = bronze, fill = bronze, below of = C2, xshift = \hd em, node distance = \vd em] (S23){$S_3$};
\draw[->] (S21.north) -- node[xshift = -1.1em, scale = 0.8] {$OK_1$}(C2.south west);
\draw[->] (S22.north) -- node[xshift = 0.8em, scale = 0.8] {$OK_2$}(C2.south);
\node[circle, draw = amaranth, fill = amaranth, scale=0.6, right of = C2, node distance = 7*\hd em] (C3){C};
\node[rectangle, draw = airforceblue, fill = airforceblue, below of = C3, xshift =  -1*\hd em, node distance = \vd em] (S31){$S_1$};
\node[rectangle, draw = asparagus, fill = asparagus, below of = C3, node distance = \vd em] (S32){$S_2$};
\node[rectangle, draw = bronze, fill = bronze, below of = C3, xshift = \hd em, node distance = \vd em] (S33){$S_3$};

\draw[->] (C3.south east) -- node[xshift = 1.3em, yshift = 0.25em, scale = 0.8] {commit \\ $tx$}(S33.north);

\end{tikzpicture}
}
\caption{OmniLedger}
\label{subfig:omniledger_cross_shard}
\end{subfigure}
\caption{{Existing works' coordination protocols. C denotes a client. $S_1, S_2$ are input shards, $S_3$ is output shard.}} \label{fig:distributed_tx}
\end{figure}
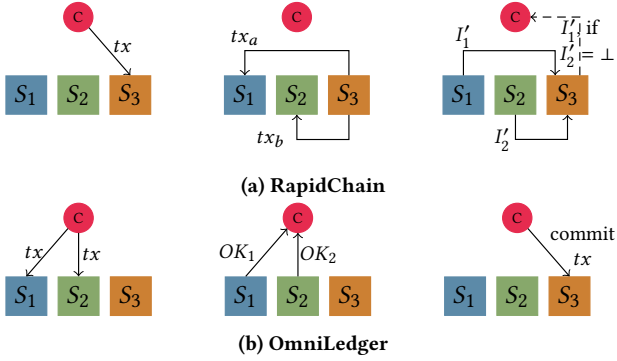
\def\hd{6}
\def\vd{2}
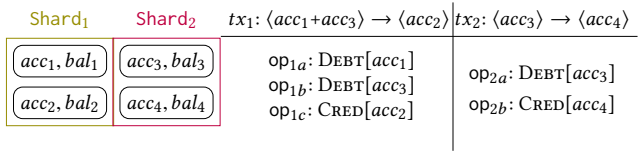
\begin{figure}
\centering    
\resizebox{.48\textwidth}{!}{  
\begin{tikzpicture}[scale=1, every text node part/.style={align=center}]  
\node[rectangle, rounded corners, draw = black] (a1){$acc_1, bal_1$};
\node[rectangle, rounded corners, draw = black, below of = a1, node distance = \vd em] (a2){$acc_2, bal_2$};
\node[rectangle, rounded corners, draw = black, right of = a1, node distance = 0.85*\hd em] (a3){$acc_3, bal_3$};
\node[rectangle, rounded corners, draw = black, below of = a3, node distance = \vd em] (a4){$acc_4, bal_4$};
\draw[olive] ($(a1.north west)+(-0.1,0.1)$)  rectangle ($(a2.south east)+(0.1,-0.1)$);
\draw[purple] ($(a3.north west)+(-0.1,0.1)$)  rectangle ($(a4.south east)+(0.1,-0.1)$);
\node[rectangle, draw = none, above of = a1, node distance = 2em] (s1){\textcolor{olive}{$\texttt{Shard}_1$}};
\node[rectangle, draw = none, above of = a3, node distance = 2em] (s2){\textcolor{purple}{$\texttt{Shard}_2$}};

\node[rectangle, draw = none, right of = s2, node distance = 1.4*\hd em] (tx1){$tx_1$: $\langle acc_1$+$acc_3 \rangle$ $\rightarrow$ $ \langle acc_2 \rangle$ };

\node[rectangle, draw = none, right of = tx1, node distance = 1.6*\hd em] (tx2){$tx_2$: $\langle acc_3 \rangle \rightarrow \langle acc_4 \rangle$};
\draw[-] ($(tx1.south west)$) -- ($(tx2.south east)$);
\draw[-] ($(tx1.north east)+(-0.15,0.)$) -- ($(tx1.north east)+(-0.15,-2.25)$);

\node[rectangle, draw = none, below of = tx1, node distance = \vd em] (op11){$\texttt{op}_{1a}$: \textproc{Debt}[$acc_1$]};
\node[rectangle, draw = none, below of = op11, node distance = 1.2 em] (op12){$\texttt{op}_{1b}$: \textproc{Debt}[$acc_3$]};
\node[rectangle, draw = none, below of = op12, node distance = 1.2 em] (op13){$\texttt{op}_{1c}$: \textproc{Cred}[$acc_2$]};

\node[rectangle, draw = none, below of = tx2, node distance = 1.3* \vd em] (op21){$\texttt{op}_{2a}$: \textproc{Debt}[$acc_3$]};
\node[rectangle, draw = none, below of = op21, node distance = 1.5 em] (op22){$\texttt{op}_{2b}$: \textproc{Cred}[$acc_4$]};

\end{tikzpicture}
}
\vspace{-3mm}
\caption{{Account-based cross-shard transactions.}}

\label{fig:example}
\end{figure}

The simplicity of UTXO model is exploited in previous works to
achieve atomicity without using a distributed commit protocol. Consider a simple
UTXO transaction $tx=\langle (I_1, I_2), O\rangle$ that spends coins $I_1, I_2$
in shard $S_1$ and $S_2$, respectively, to create a new coin $O$ belonging to
shard $S_3$ (Figure~\ref{subfig:Rapidchain_cross_shard}).
RapidChain~\cite{rapidchain} executes $tx$ by splitting it into three
sub-transactions:  $tx_a=\langle I_1, I'_1\rangle$, $tx_b=\langle I_2,
I'_2\rangle$, $tx_c=\langle (I'_1, I'_2), O\rangle$, where $I'_1$ and $I'_2$
belong to $S_3$. $tx_a$ and $tx_b$ essentially transfer $I_1$ and $I_2$ to the
output shard, which are spent by $tx_c$ to create the final output $O$. All
three sub-transactions are single-shard. In case of failures, when, for example,
$tx_b$ fails while $tx_a$ succeeds, RapidChain sidesteps atomicity by informing
the owner of $I_1$ to use $I_1'$ for future transactions, which has the same
effect as rolling back the failed $tx$.

RapidChain does not achieve isolation.  Consider another transaction $tx_b'$ in $S_2$ that spends $I_2$ and is
submitted roughly at the same time as $tx$, the shard serializes the transactions, thus only one of $tx_b$ and
$tx_b'$ succeeds. If isolation is achieved, either $tx$ or $tx_b'$ succeeds. But it is possible in RapidChain
that both of them fail, because $tx_a$ fails. 

\vspace{2mm}
{\bf Safety for general transaction model.}
We now show examples demonstrating how RapidChain's approach fails to work for
non-UTXO distributed transactions, because it violates both atomicity and
isolation. Consider the account-based data model, which is used in Ethereum. Let
$tx_1$: $\langle acc_1$+$acc_3 \rangle$ $\rightarrow$ $ \langle acc_2 \rangle$
be a transaction transferring assets from $acc_1$ and $acc_3$ to $acc_2$, where
$acc_1, acc_2$ belongs to shard $S_1$ and $acc_3$ belongs to shard $S_2$.
Following RapidChain, $tx_1$ is split into $op_{1a}, op_{1b}, op_{1c}$
(Figure~\ref{fig:example}). If $op_{1a}$ succeeds and $op_{1b}$ fails, due to
insufficient funds, for example, $op_{1c}$ cannot be executed. In other words,
$tx_1$ does not achieve atomicity because it is executed only partially: $acc_1$
is already debited and cannot be rolled back.

Let $tx_2$: $\langle acc_3 \rangle \rightarrow \langle acc_4 \rangle$ be another transaction
submitted roughly at the same time as $tx_1$. In Figure~\ref{fig:example}, the execution sequence $\langle
op_{1a}, op_{1b}, op_{2a}, op_{2b}, op_{1c}\rangle$ is valid in RapidChain, but it breaks isolation
(serializability) because $tx_2$ sees the states of a partially completed transaction.

\vspace{2mm}
{\bf Liveness under malicious coordinator.}
OmniLedger~\cite{omniledger} achieves safety for the UTXO model by relying on a
client to coordinate a \emph{lock/unlock} protocol
(\autoref{subfig:omniledger_cross_shard}). Given a transaction $tx$ whose inputs
belong to shard $S_1$ and $S_2$, and output belongs to shard $S_3$, the client
first obtains locks from $S_1$ and $S_2$ (i.e., marking the inputs as spent),
before instructing $S_3$ to commit $tx$.  This client-driven protocol suffers
from indefinite blocking if the client is malicious. For example, consider a
payment channel~\cite{lightning, sprites}, in which the payee is the client that
coordinates a transaction that transfers funds from a payer's account. A
malicious payee may pretend to crash indefinitely during the lock/unlock
protocol, hence, the payer's funds are locked forever.

\subsection{Our Solution}
\label{subsec:cross_shard_atomicity}
\begin{figure}
\centering
\includegraphics[width=0.4\textwidth]{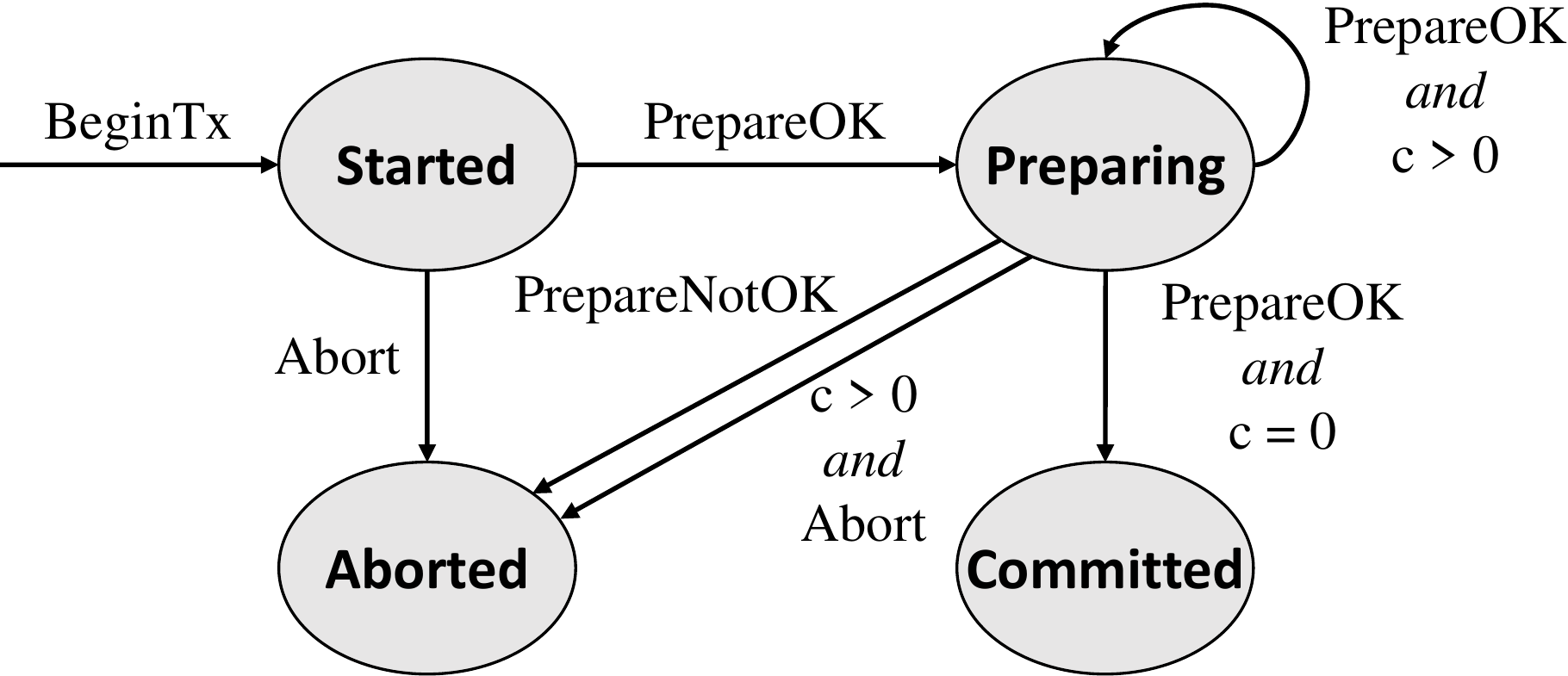}
\caption{State machine of the reference committee.}
\label{fig:statemachine}
\end{figure}

{ We aim to design a distributed transaction protocol that
achieves safety for general blockchain transactions (non-UTXO), and liveness
against malicious coordinators. For safety, we use the classic two-phase commit
(2PC) and two-phase locking (2PL) protocol as in traditional databases. To guard
against a malicious coordinator, we employ a Byzantine fault-tolerant reference
committee, denoted by $R$, to serve as a coordinator. $R$ runs BFT consensus
protocol and implements a simple state machine for 2PC. Given our system and
threat model in Section~\ref{subsec:threat_model}, $R$ is highly (eventually)
available.  \autoref{fig:ours_cross_shard} illustrates the transaction flow and
\autoref{fig:statemachine} depicts the state machine of the reference
committee.}

\def\hd{2}
\def\vd{3}
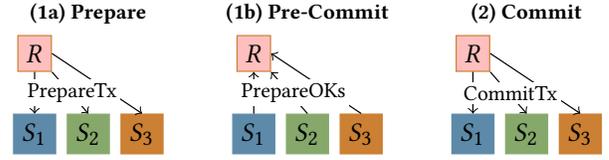
\begin{figure}
\centering    
\resizebox{.45\textwidth}{!}{  
\begin{tikzpicture}[scale=0.6, every text node part/.style={align=center}]  

\node[circle, draw = none, fill = none, scale=0.68] (C2){};
\node[rectangle, draw = bronze, fill = pink, left of = C2, node distance = 2*\hd em] (CS2){$R$};
\node[rectangle, draw = airforceblue, fill = airforceblue, below of = C2, xshift = -2*\hd em, node distance = \vd em] (S21){$S_1$};
\node[rectangle, draw = asparagus, fill = asparagus, below of = C2, xshift = -1*\hd em, node distance = \vd em] (S22){$S_2$};
\node[rectangle, draw = bronze, fill = bronze, below of = C2,  node distance = \vd em] (S23){$S_3$};
\draw[->] (CS2.east) -- (S23.north);
\draw[->] (CS2.south) -- (S21.north);
\draw[->] (CS2.south east) -- node[scale = 0.8, xshift = 0.1em, fill = white, inner sep =1.5pt] {PrepareTx}(S22.north);

\node[circle, draw = none, fill = none, scale=0.68, right of = C2, node distance = 6*\hd em] (C3){};
\node[rectangle, draw = airforceblue, fill = airforceblue, below of = C3, xshift = -2*\hd em, node distance = \vd em] (S31){$S_1$};
\node[rectangle, draw = asparagus, fill = asparagus, below of = C3, xshift = -1*\hd em, node distance = \vd em] (S32){$S_2$};
\node[rectangle, draw = bronze, fill = bronze, below of = C3,  node distance = \vd em] (S33){$S_3$};
\node[rectangle, draw = bronze, fill = pink, left of = C3, node distance = 2*\hd em] (CS3){$R$};
\draw[<-] (CS3.east) -- (S33.north);
\draw[<-] (CS3.south) -- (S31.north);
\draw[<-] (CS3.south east) -- node[scale = 0.8, xshift = 0.2em, fill = white, inner sep =1.5pt] {PrepareOKs}(S32.north);

\node[circle, draw = none, fill = none, scale=0.68, right of = C3, node distance = 6*\hd em] (C4){};
\node[rectangle, draw = airforceblue, fill = airforceblue, below of = C4, xshift = -2*\hd em, node distance = \vd em] (S41){$S_1$};
\node[rectangle, draw = asparagus, fill = asparagus, below of = C4, xshift = -1*\hd em, node distance = \vd em] (S42){$S_2$};
\node[rectangle, draw = bronze, fill = bronze, below of = C4,  node distance = \vd em] (S43){$S_3$};
\node[rectangle, draw = bronze, fill = pink, left of = C4, node distance = 2*\hd em] (CS4){$R$};
\draw[->] (CS4.east) -- (S43.north);
\draw[->] (CS4.south) -- (S41.north);
\draw[->] (CS4.south east) -- node[scale = 0.8, xshift = 0.1em, fill = white, inner sep =1.5pt] {CommitTx}(S42.north);

\node[rectangle, draw = none, above of = S22, node distance = 1.5*\vd em, scale = 0.8]{\textbf{(1a) Prepare}};
\node[rectangle, draw = none, above of = S32, node distance = 1.5*\vd em, scale = 0.8]{\textbf{(1b) Pre-Commit}};
\node[rectangle, draw = none, above of = S42, node distance = 1.5*\vd em, scale = 0.8]{\textbf{(2) Commit}};

\end{tikzpicture}
}
\caption{Our coordination protocol.}
\label{fig:ours_cross_shard}
\end{figure}

{
The client initiates a transaction $tx$ by sending \texttt{BeginTx} request to the reference committee.
The transaction then proceeds in three steps.}

\vspace{2mm}
\noindent\textbf{1a) Prepare}:
{Once $R$ has executed the \texttt{BeginTx} request, it enters
Started state.  Nodes in $R$ then send \texttt{PrepareTx} requests to the
transaction committees (or tx-committees). The latter wait for a quorum of
matching \texttt{PrepareTx} to ensure that \texttt{BeginTx} has been executed in
$R$.  Each tx-committee executes the \texttt{PrepareTx}. If consensus is reached
that $tx$ can be committed, which requires that $tx$ can obtain all of its
locks, the nodes within the committee send out \texttt{PrepareOK} messages (or
\texttt{PrepareNotOK} messages otherwise).}

\vspace{2mm}
\noindent\textbf{1b) Pre-Commit}: 
{
When entering Started state, $R$ initializes a counter $c$ with the number of tx-committees involved in $tx$.
After receiving the first quorum of matching responses from a tx-committee, it either decreases $c$ and enters
the Preparing state, or enters the Aborted state, depending on whether the responses are \texttt{PrepareOK} or
\texttt{PrepareNotOK} respectively. $R$ stays in the Preparing states and decreases $c$ for every new quorum
of \texttt{PrepareOK} responses. It moves to Aborted as soon as it receives a quorum of \texttt{PrepareNotOK},
and to Committed states when $c=0$.} 


\vspace{2mm}
\noindent\textbf{2) Commit}: 
{%
Once $R$ has entered Committed (or Aborted) state, the nodes in $R$ send out \texttt{CommitTx} (or
\texttt{AbortTx}) messages to tx-committees. The latter wait for a quorum of matching messages from $R$ before
executing the corresponding commit or abort operation.}

\vspace{2mm}
{ We remark  that the reference committee is not a bottle-neck in cross-shard transaction
processing, for we can scale it out by running multiple instances of $R$ in parallel.  
}

\begin{figure}
  \centering
  \includegraphics[width=.85\linewidth]{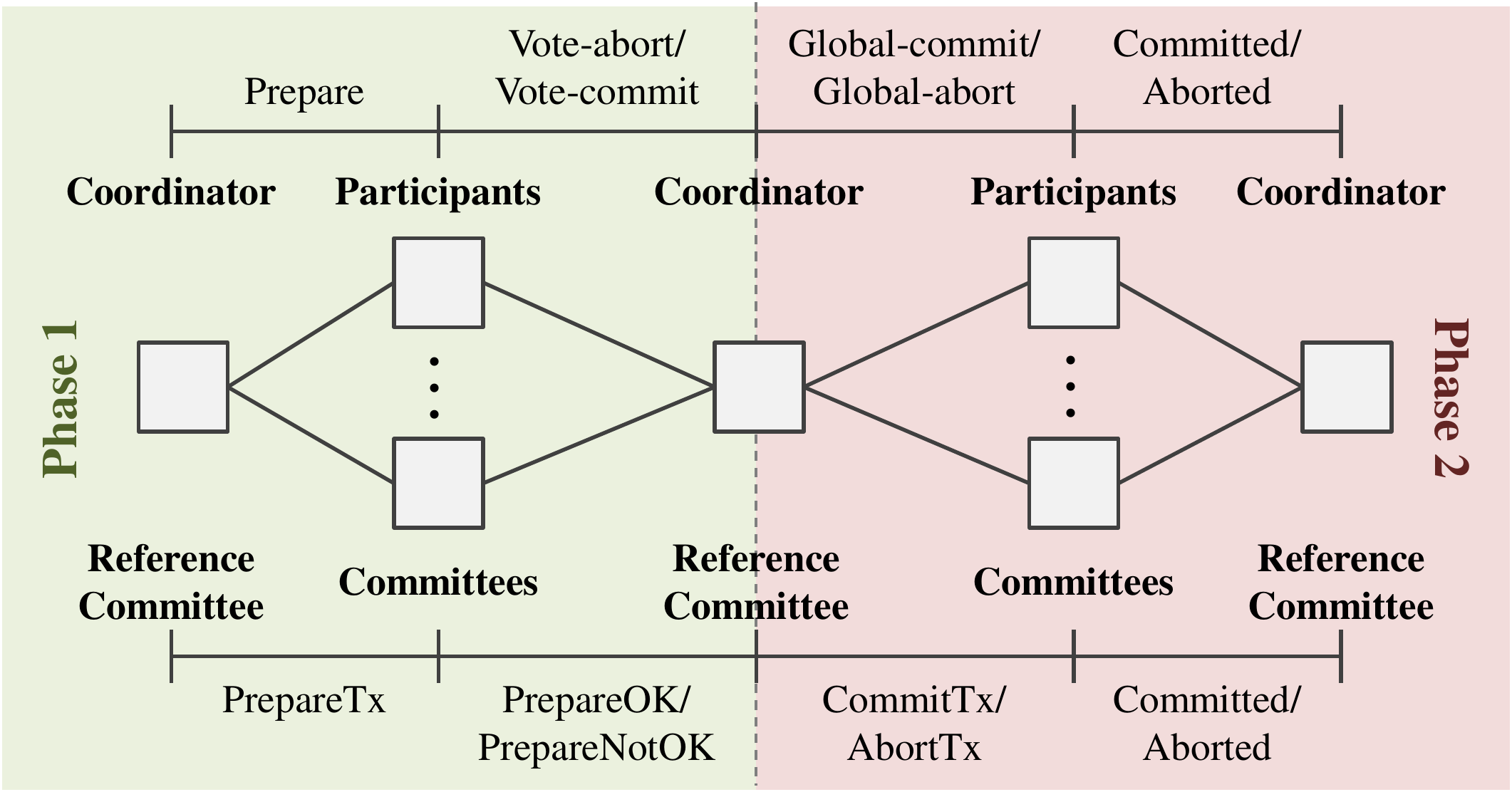}
  \caption{{Correspondence between our distributed transaction management protocol (i.e., bottom half) and the original 2PC protocol (i.e., top half).}}
  \label{fig:2pc_proof}
\end{figure}

\vspace{2mm}
\noindent\textbf{Safety analysis.} 
{ The safety of our coordination protocol is based on the assumption that both $R$ and
tx-committees ensure safety for all transactions/requests they process. This assumption is realized by
fine-tuning the committee size according to Equation~\ref{eq:comm_fault}
presented in Section~\ref{subsec:comm_size}.  }

{
We sketch the proof that our coordination protocol indeed implements the classic 2PC protocol in which reference
committee $R$ is the \textit{coordinator}, and tx-committees are the participants. The state machine of the
reference committee shown in \autoref{fig:statemachine} is identical to that of the coordinator in the
original 2PC~\cite{2PC}. }

{ \autoref{fig:2pc_proof} illustrates the correspondence between
our protocol and the original 2PC protocol. Similar to 2PC, our protocol
consists of two main phases. Phase 1 aims to reach the tentative agreement of
transaction commit and Phase 2 performs the actual commit of the transaction
among shards. Before \texttt{BeginTx} is executed at $R$, the transaction is
considered non-existent, hence no tx-committees would accept it.  After $R$
enters Started state (i.e., it has logged the transaction), the
\texttt{PrepareTx} requests are sent to tx-committees. Phase 1 completes when
$R$ moves either to Committed or Aborted state. At this point, the current state
of $R$ reflects the tentative agreement of transaction commit.
When this tentative agreement is conveyed to the tx-committees in Phase 2, they
can commit  (or abort) the transaction. The original 2PC requires logging at the
coordinator and participants for recovery. Our protocol, however, does not need
such logging, because the states of $R$ and of tx-committees are already stored
on the blockchain. In summary, our protocol always achieves safety for
distributed transactions.}



\vspace{2mm}
\noindent\textbf{Liveness analysis.}
{
Recall that we assume a partially synchronous network, in which messages sent
repeatedly with a finite time-out will eventually be received. Furthermore, we assume that the size of $R$ is chosen such that
the number of Byzantine nodes are less than half. Under these assumptions, the BFT protocol running in $R$
achieves liveness. In other words, $R$ always makes progress, and any request sent to it will eventually be
processed. Such eventual availability means that $R$ will not block indefinitely. Thus, the coordination
protocol achieves liveness. }


\subsection{Implementation}
We implement our protocol on Hyperledger Fabric which supports smart contracts called chaincodes. The
blockchain states are modeled as key-value tuples and accessible to the chaincode during execution.
{
We use the chaincode that implements the SmallBank benchmark to explain our implementation. In Hyperledger,
this chaincode contains a {\tt sendPayment} function that reads the state representing $acc_1$'s balance,
checks that it is greater than $bal$, then deducts the $bal$ from $acc_1$ and updates the state representing
$acc_2$'s balance.  This chaincode does not support sharding, because the states of $acc_1$ and $acc_2$ may
belong to different shards. We modify the chaincode so that it can work with our protocol. In particular, for
the {\tt sendPayment} function, we split it into three functions: {\tt preparePayment}, {\tt commitPayment},
and {\tt abortPayment}. We implement locking for an account $acc$ by storing a boolean value to a blockchain
state with the key $``L\_"acc$. During the execution of {\tt preparePayment},
the chaincode checks if the corresponding lock, namely the tuple $\langle L\_{acc}, \textit{true}\rangle$, exists in the blockchain state,
and aborts the transaction if it does. If it does not, the chaincode writes the lock to the blockchain. The {\tt
commitPayment} function for a transaction $tx$ writes new states (balances) to the blockchain, and removes the
locks that were written for $tx$.} 

{
As an optimization to avoid cross-shard communication in normal case (when clients are honest), we let the
clients collect and relay messages between $R$ and tx-committees.  We directly exploit the blockchain's ledger
to record the progress of the commit protocol.  In particular, during Prepare phase, the client sends a
transaction to the blockchain that invokes the {\tt preparePayment} function. This function returns an error
if the Prepare phase fails. The client reads the status of this transaction from the blocks to determines if
  the result is PrepareOK or PrepareNotOK. We implement the state machine of the reference committee as a
  chaincode with similar functions that can be invoked during the two phases of our protocol. When interacting
  with $R$, all transactions are successful, therefore the client only needs to wait for them to appear on the
  blocks of $R$.    
}

\subsection{Discussion}
Our current design uses 2PL for concurrency control, which may not be able to
extract sufficient concurrency from the workload. State-of-the-art concurrency
control protocols have demonstrated superior performance over 2PL~\cite{janus,carousel}. We note that
the batching nature of blockchain presents opportunities for optimizing
concurrency control protocols. We leave the study of these protocols to future
work.

In the current implementation, we manually refactor existing chaincodes to
support sharding. One immediate extension that makes it easier to port legacy
blockchain applications to our system is to instrument Hyperledger codebase with
a library containing common functionalities for sharded applications. One common
function is state locking. Having such a library helps speed up the refactoring,
but the developer still needs to split the original chaincode function to
smaller functions that process the Prepare, Commit or Abort requests. Therefore,
a more useful extension is to add programming language features that, given a
single-shard chaincode implementation, automatically analyze the functions and
transform them to support multi-shards execution. Another extension to improve
usability is to introduce a client library that hides the details of the
coordination protocols, so that the users only see single-shard transactions.


\section{Performance Evaluation}
\label{sec:eval}

In this section, we present a comprehensive evaluation of our design. We first
demonstrate the performance of the scalable consensus protocols. Next, we report the efficiency of our shard
formation protocol. Finally, we evaluate the scalability of our sharding approach. Table~\ref{tab:comparison}
contrasts our design and evaluation methodology against existing sharded blockchain systems.

\begin{table}[]
\centering
\caption{Comparisons with other sharded blockchains.} 
\footnotesize
\resizebox{0.49\textwidth}{!}{
\begin{tabular}{|>{\centering\arraybackslash}m{1.8cm}|>{\centering\arraybackslash}m{1.5cm}|>{\centering\arraybackslash}m{1.65cm}|>{\centering\arraybackslash}m{1.5cm}|>{\centering\arraybackslash}m{1.5cm}|>{\centering\arraybackslash}m{1.5cm}|}

\hline
           & \# machines & Over-subscription & Transaction model & Distributed transaction \\ \hline
Elastico   & 800               & 2              & UTXO              & $\times$                                        \\ \hline
OmniLedger & 60                & 67             & UTXO              & $\times$                                       \\ \hline
RapidChain & 32                & 125            & UTXO              &
$\checkmark$                    \\ \hline Ours       & 1400                   & 1              & General workload &
$\checkmark$          				 \\ \hline
\end{tabular}
}
\label{tab:comparison}
\end{table}

For this evaluation, we use KVStore and Smallbank, two different benchmarks
available in BLOCKBENCH~\cite{blockbench} --- a framework for benchmarking
private blockchains. We use the original client driver in BLOCKBENCH, which is
open-loop, for our single-shard experiments. For multi-shard experiments, we
modified the driver to be closed-loop (i.e., it waits until a cross-shard
transaction finishes before issuing a new one). To generate cross-shard
transactions, we modified the original KVStore driver to issue 3 updates per
transaction, and used the original {\tt sendPayment} transaction in Smallbank
that reads and writes two different states.

We conducted experiments in two different settings. One is an in-house (local)
cluster consisting of $100$ servers, each equipped with Intel Xeon E5-1650
3.5GHz CPUs, 32GB RAM and 2TB hard drive. In this setting, the blockchain node
and client run on separate servers. The other setting is Google Cloud Platform
(GCP), in which we have separate instances for the clients and for the nodes. A
client has 16 vCPUs and 32GB RAM, while a node has 2 vCPUs and 12GB RAM. We use
up to $1400$ instances over 8 regions (the latency between regions is included
in Appendix~\ref{sec:apd_experiments}).


\begin{table}[t]
\centering
\caption{{Runtime costs of enclave operations (excluding enclave switching cost which is roughly $2.7\mu s$).}}
\label{tab:sgx_runtime}
\resizebox{.9\linewidth}{!}{
\begin{tabular}{>{\raggedright\arraybackslash}m{5.25cm}>{\raggedleft\arraybackslash}m{3.25cm}}
\hline
\textbf{Operations} & \textbf{Time $(\mu s)$}         \\
\hline

ECDSA Signing       & $458.4(\pm0.4) $  \\
ECDSA Verfication   & $844.2(\pm0.8) $  \\
SHA256			    & $2.5(\pm0.1) $  \\
AHL Append          & $465.3 (\pm 0.8) $   \\
AHLR Message Aggregation ($f = 8)$          & $8031.2 (\pm2.3) $  \\
\textproc{RandomnessBeacon}           & $482.2(\pm0.5) $  \\
\hline
\end{tabular}
}
\end{table}

We used Intel SGX SDK~\cite{sgx_sdk} to implement the trusted code base. Since
SGX is not available on the local cluster and GCP, we configured the SDK to run
in simulation mode.
{ We measured the latency of each SGX operation on Skylake
6970HQ 2.80 GHz CPU with SGX Enabled BIOS support, and injected it to the
simulation. Table \ref{tab:sgx_runtime} details runtime costs of enclave
operations on the SGX-enabled processor. Public key operations are expensive:
signing and signature verification take about $450\mu s$ and $844 \mu s$,
respectively. Context switching and other symmetric key operations take less
than $5 \mu s$. We also measured the cost of remote attestation protocol, which
is carried out between nodes of the same committee in order to verify that they
are running the correct enclave.  On our SGX-enabled processor, this protocol
takes around $2ms$, but we note that it is executed only once per epoch, and its
results can be cached.}


The results reported in the following are averaged over ten independent runs. Due to space
constraints, we focus on throughput performance in this section, and discuss other results in the Appendix.

\begin{figure}
\includegraphics[width=0.47\textwidth]{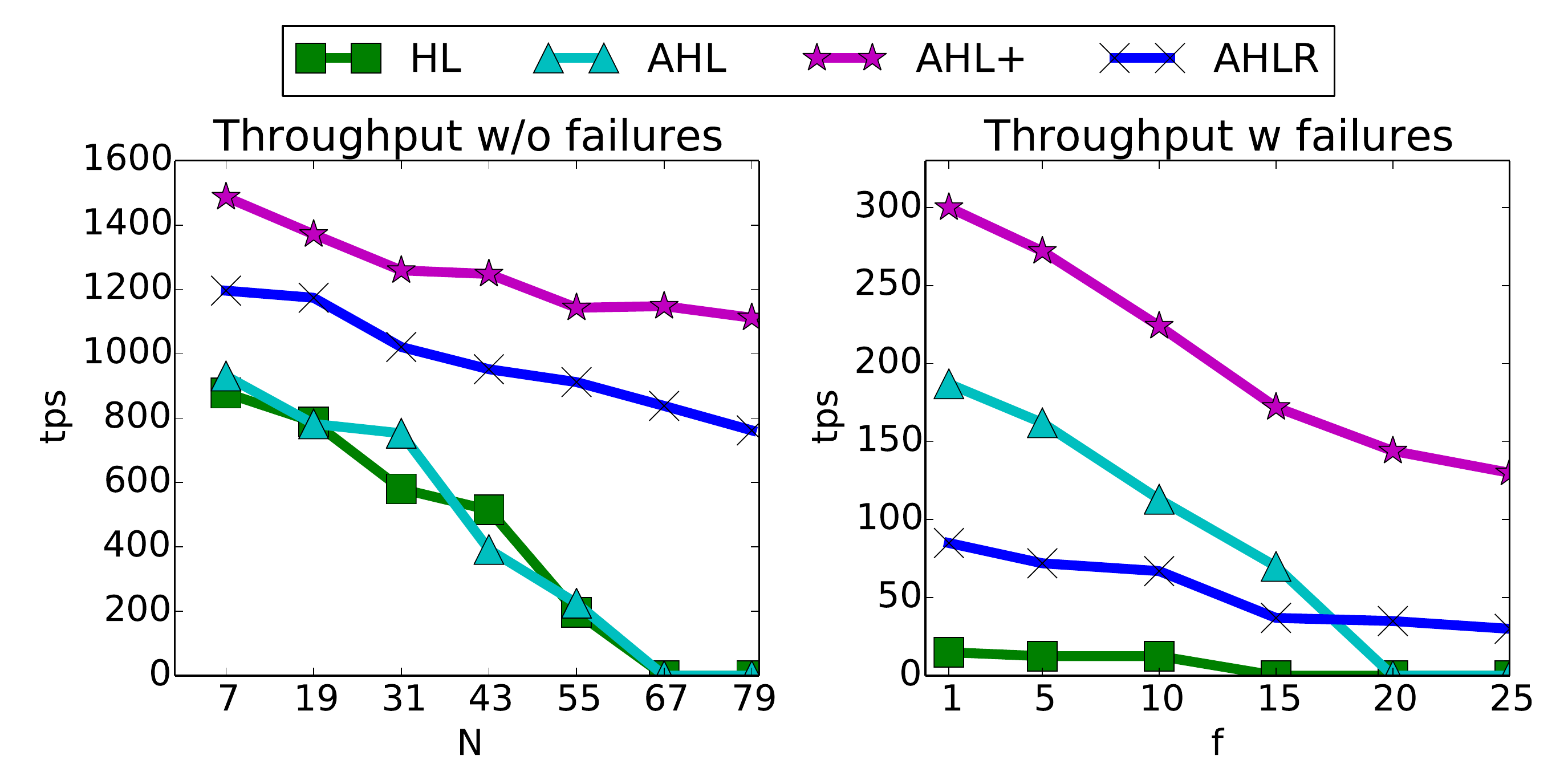}
\caption{AHL+ performance on local cluster.}
\label{fig:cluster_bft_f}
\end{figure}

\begin{figure}
\includegraphics[width=0.47\textwidth]{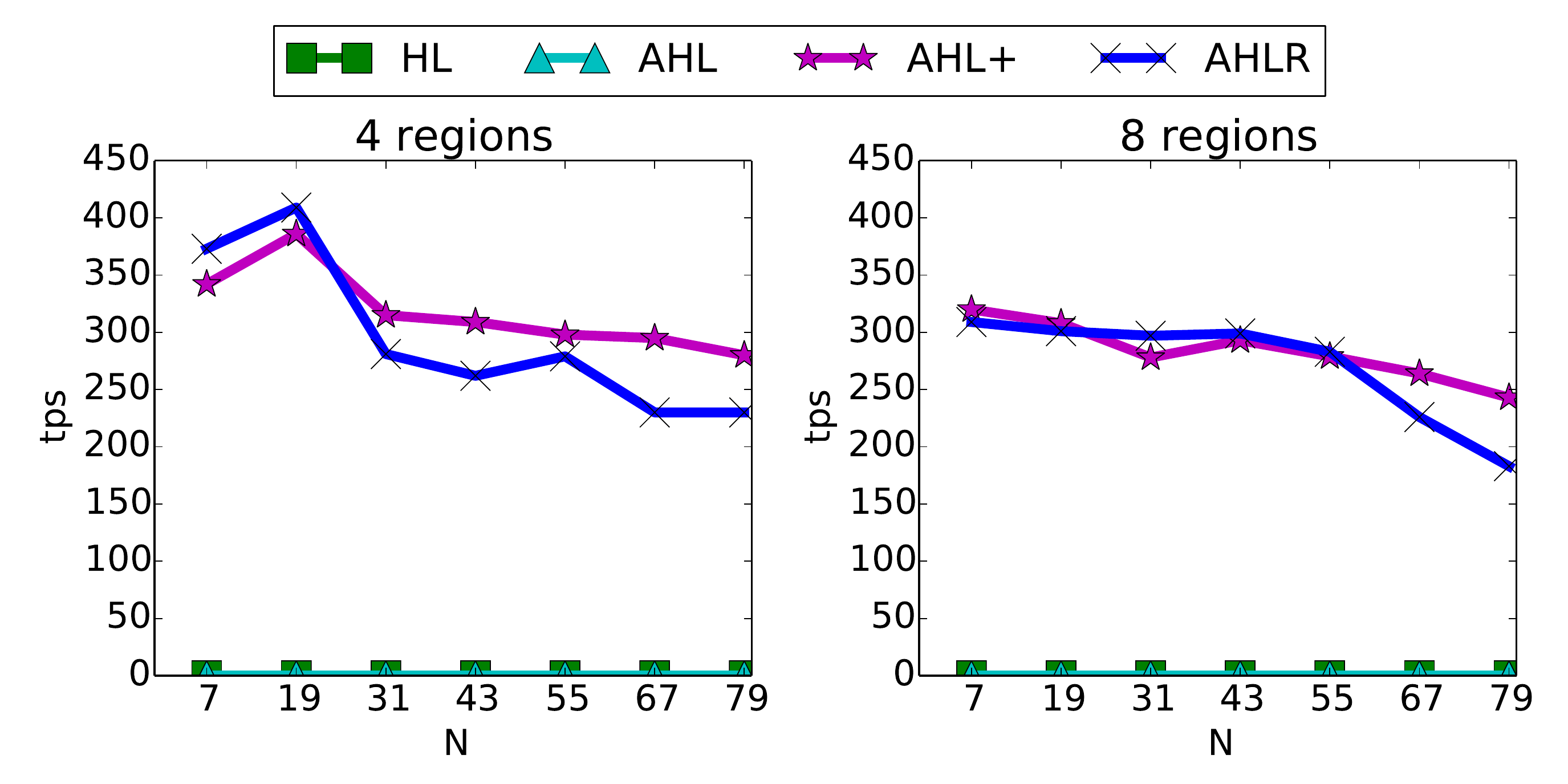}
\caption{AHL+ performance on GCP (4 and 8 regions).}
\label{fig:gcp_bft_f}
\end{figure}

\begin{figure}
\centering
\includegraphics[width=0.47\textwidth]{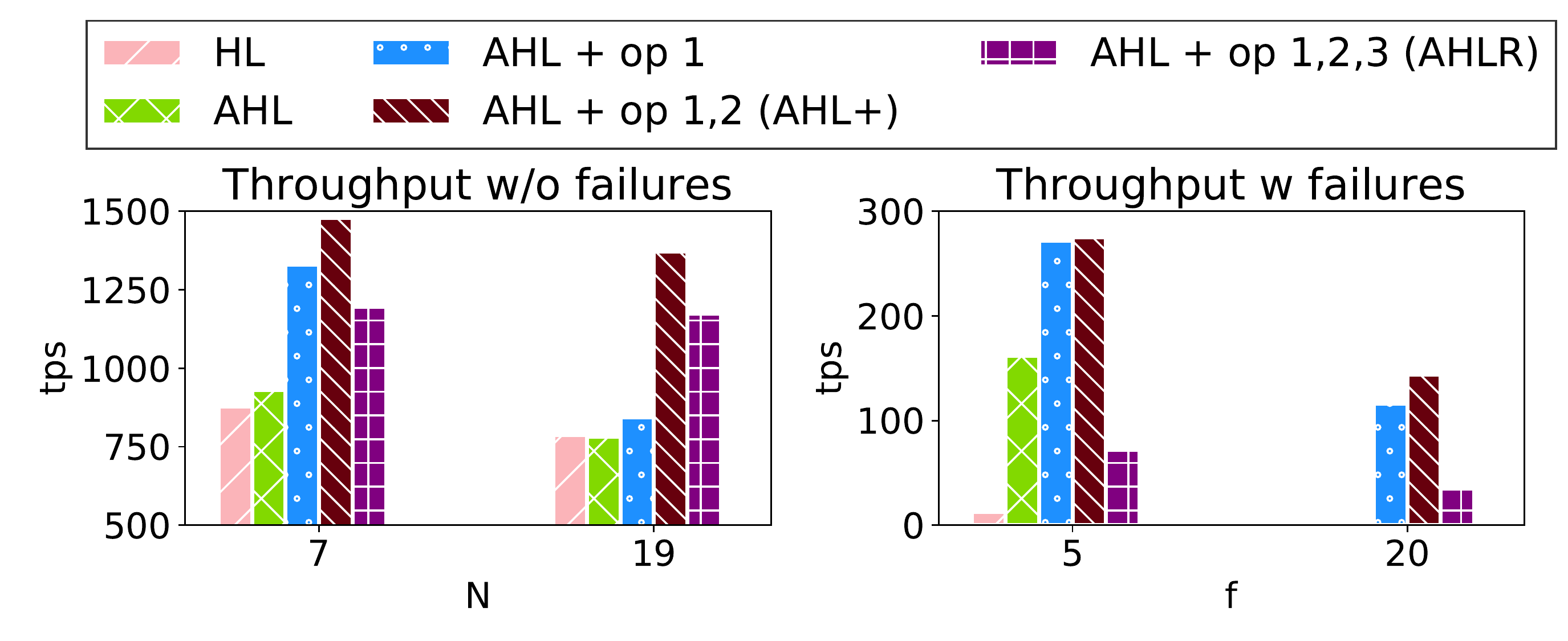}
\caption{{Effect of optimizations on throughput.}}
\label{fig:optimizations}
\end{figure}

\subsection{Fault-scalable consensus}
\label{subsec:single_shard_eval}
\noindent \textbf{AHL+ vs. other variants.} We compare the performance of AHL+
with the original PBFT protocol (denoted by HL), AHL and AHLR.
Figure~\ref{fig:cluster_bft_f} and Figure~\ref{fig:gcp_bft_f} show the
throughput with increasing number of nodes, $N$, on the local cluster and on
GCP, when using KVStore benchmark with $10$ clients. The performance with
varying number of clients and fixed $N$ is included in the Appendix.

AHL's throughput is similar to that of HL, but for the same $N$ it tolerates
more failures. Both HL and AHL show no throughput for $N>67$ on the local
cluster, and no throughput at all on GCP. We observe that these systems are
livelocked when $N$ increases, as they are stuck in the view-change phase. The
number of view changes is reported in the Appendix. In contrast, both AHL+ and
AHLR maintain throughputs above $700$ transactions per second in the cluster and
above $200$ on GCP.  Interestingly, AHL+ demonstrates consistently higher
throughput than AHLR, even though the former has $O(N^2)$ communication overhead
compared to $O(N)$ of the latter. Careful analysis of AHLR  shows that the
leader becomes the single point of failure. If the leader fails to collect and
multicast the aggregate message before the time out, the system triggers the
view change protocol which is expensive.


To understand the impact of Byzantine behavior on the overall performance, we
simulate an attack in which the Byzantine nodes send conflicting messages (with
different sequence numbers) to different nodes.
Figure~\ref{fig:cluster_bft_f} (right) shows how the throughput deteriorates
when the number of failures increases. We note that for a given $f$, HL runs
with $N=3f+1$ nodes, whereas AHL, AHL+ and AHLR run with $N=2f+1$ nodes. Despite
the lower throughputs than without failures, we observe a similar trend in which
AHL+ outperforms the other protocols. On GCP with more than 1 zone, the
Byzantine behavior causes all protocols to livelock.

{
Finally, we examine how each optimization presented in Section~\ref{subsec:ahl}
contributes to the final performance. Figure~\ref{fig:optimizations} shows,
against the baseline of original PBFT implementation (HL), the effect of adding
trusted hardware (AHL), optimization 1 (separating message queues), optimization
2 (removing request), and optimization 3 (message aggregation at the leader).
The experiments, which are run on our local cluster with $10$ clients, show that
optimization 2 adds the most benefits when there is no failure, whereas
optimization 1 is the best under Byzantine failures. This explains why AHL+,
which incorporates both optimizations, has the best performance.}

\subsection{Shard Formation}
\label{subsec:shard_formation_eval}
\begin{figure}
\centering
\includegraphics[width=0.45\textwidth]{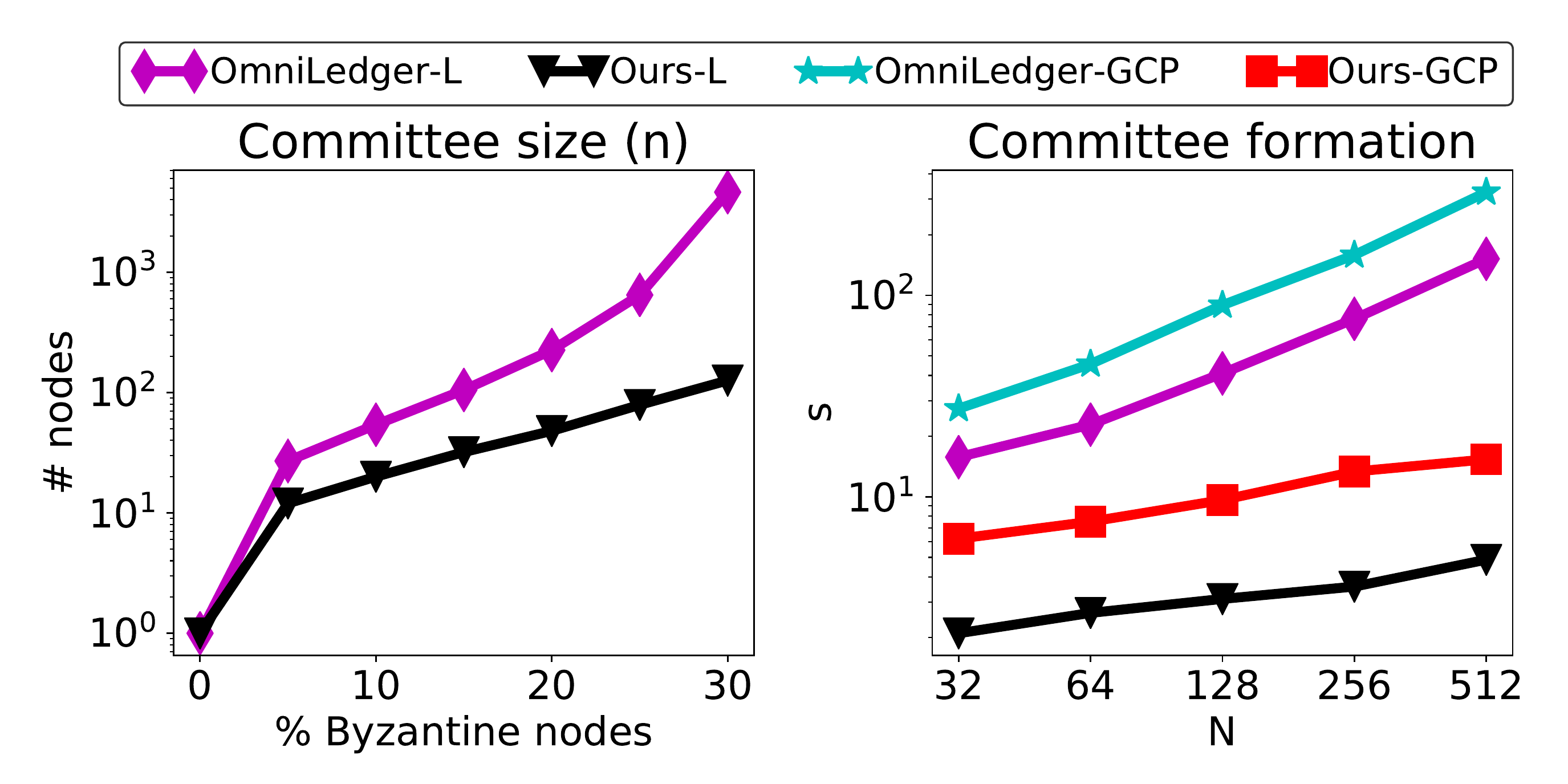}
\caption{Evaluation of shard formation.}
\label{fig:shard_formation}
\end{figure}

\begin{figure}
\centering
\includegraphics[width=0.49\textwidth]{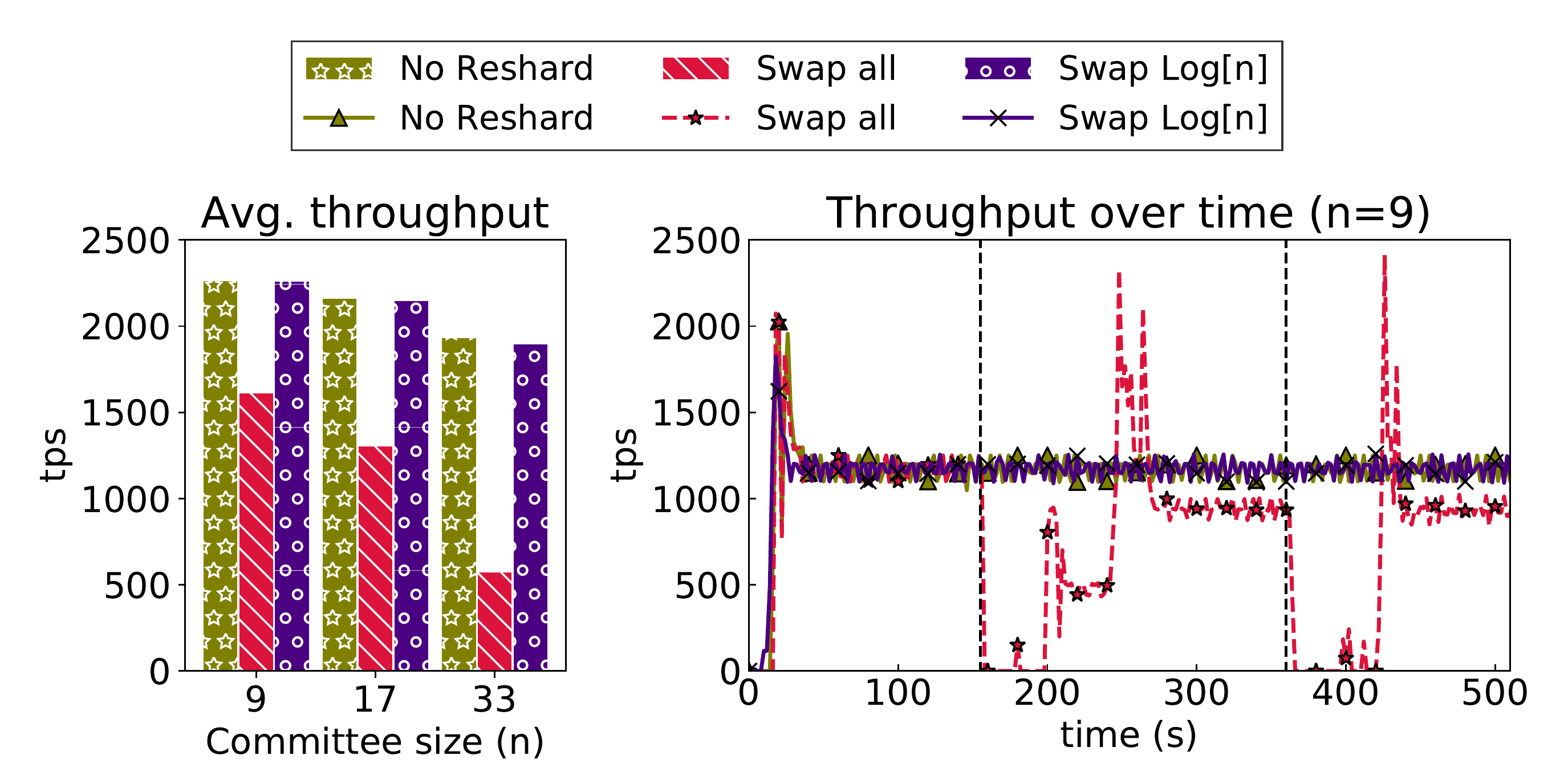}
\caption{{Performance during shard reconfiguration.}}
\label{fig:epoch_trans_tps}
\end{figure}


Figure ~\ref{fig:shard_formation} compares our shard formation protocol with
OmniLedger's in terms of committee size and running time. With increasing
Byzantine failures, OmniLedger needs exponentially larger committees. On the
other hand, by leveraging AHL+, our protocol keeps the committees up to two
orders of magnitude smaller.

We compare the cost of our distributed random number generation protocol with
that of Randhound used in OmniLedger. We vary the network size from $32$ to
$512$ nodes. On the local cluster, we oversubscribe each physical server by a
factor of $8$, running each node in a single-threaded virtual machine.  On GCP,
each node runs on an instance with 2 vCPUs. Recall that both protocols assume a
synchronous network with the bounded delay $\Delta$. We empirically derive
$\Delta$ by measuring the maximum propagation delay in different network sizes
for a 1KB message, then conservatively setting $\Delta$ to be $3\times$ the
measured value. On the local cluster, $\Delta$ ranges from $2$ to $4.5$s. On
GCP, $\Delta$ ranges from $5.9$ to $15$s. We set $q = \log(N) - \log(\log(N))$,
so that the communication overhead is $O(N\log(N))$ and $P_{repeat} < 2^{-11}$.
For Randhound, we set $c=16$ as suggested in~\cite{omniledger}.
Figure~\ref{fig:shard_formation} shows that our protocol is up to 32$\times$ and
21$\times$ faster than RandHound on the local cluster and GCP, respectively. We
attribute this gap to the difference in their communication complexity:
$O(N\log(N))$ versus $O(Nc^2)$.

{
\vspace{2mm}
\noindent{\bf Shard Reconfiguration.}
Next, we analyze the performance of our system during epoch
transition (or resharding).  We consider the naive approach which \textit{swaps all} nodes by first stopping 
them, assigning them to new shards based on a randomly generated permutation, and then starting them again.
We compare it with our approach which swaps \texttt{B} nodes at a time. In our evaluation, we set $\texttt{B}
= \log(n)$.

Figure~\ref{fig:epoch_trans_tps} shows the throughput against the baseline where there is no
resharding. We run the experiments on our local clusters with two shards, each containing up to $33$ nodes.
We perform the resharding twice, as depicted in Figure~\ref{fig:epoch_trans_tps}~(right).  The naive
approach shows a sharp drop in throughput when all nodes are restarted, followed by a period of up to $80$s in
which nodes discover their new peers, verify and synchronize their states. The subsequent spikes in throughput
are due to nodes processing transactions from their backlog.  In contrast, our approach allows the system to
maintain the same throughput as the baseline.
}  

\begin{figure}
\centering
\begin{subfigure}{0.235\textwidth}
\includegraphics[width=1\textwidth]{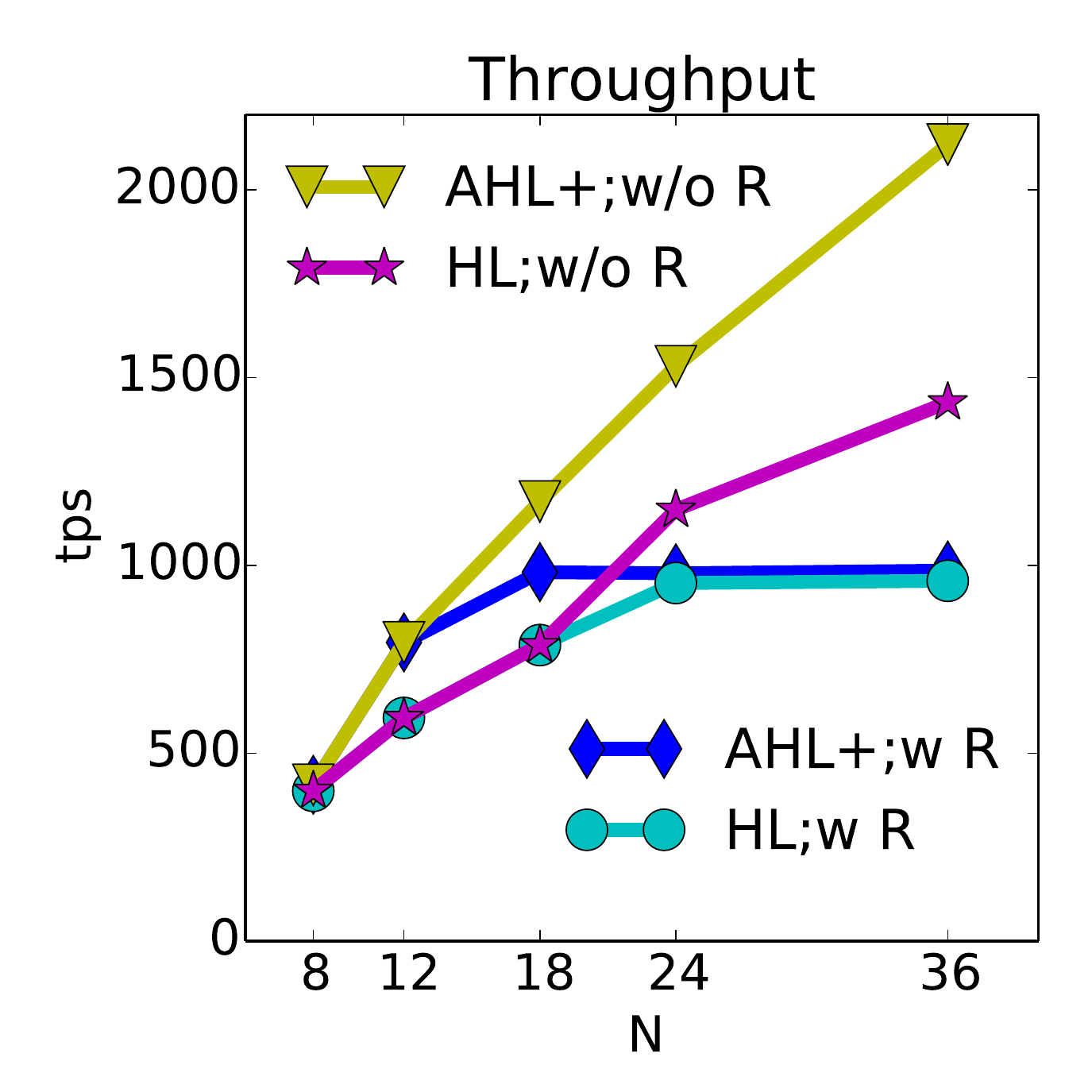}
\end{subfigure}
\hfill
\begin{subfigure}{0.235\textwidth}
\includegraphics[width=1\textwidth]{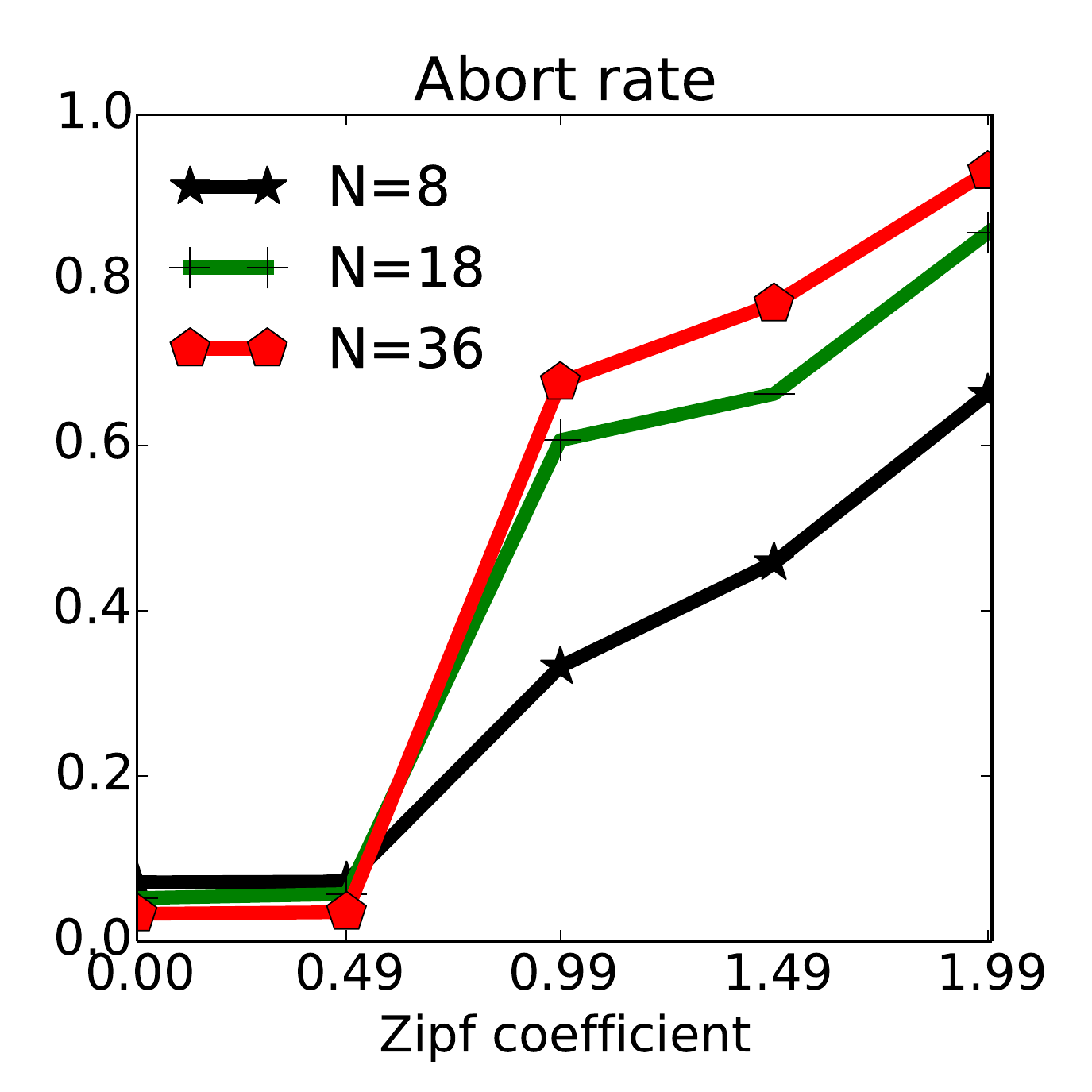}
\end{subfigure}
\caption{Sharding performance on local cluster with and without reference committee.}
\label{fig:cluster_sharding}
\end{figure}

\begin{figure}
\centering
\includegraphics[width=0.45\textwidth]{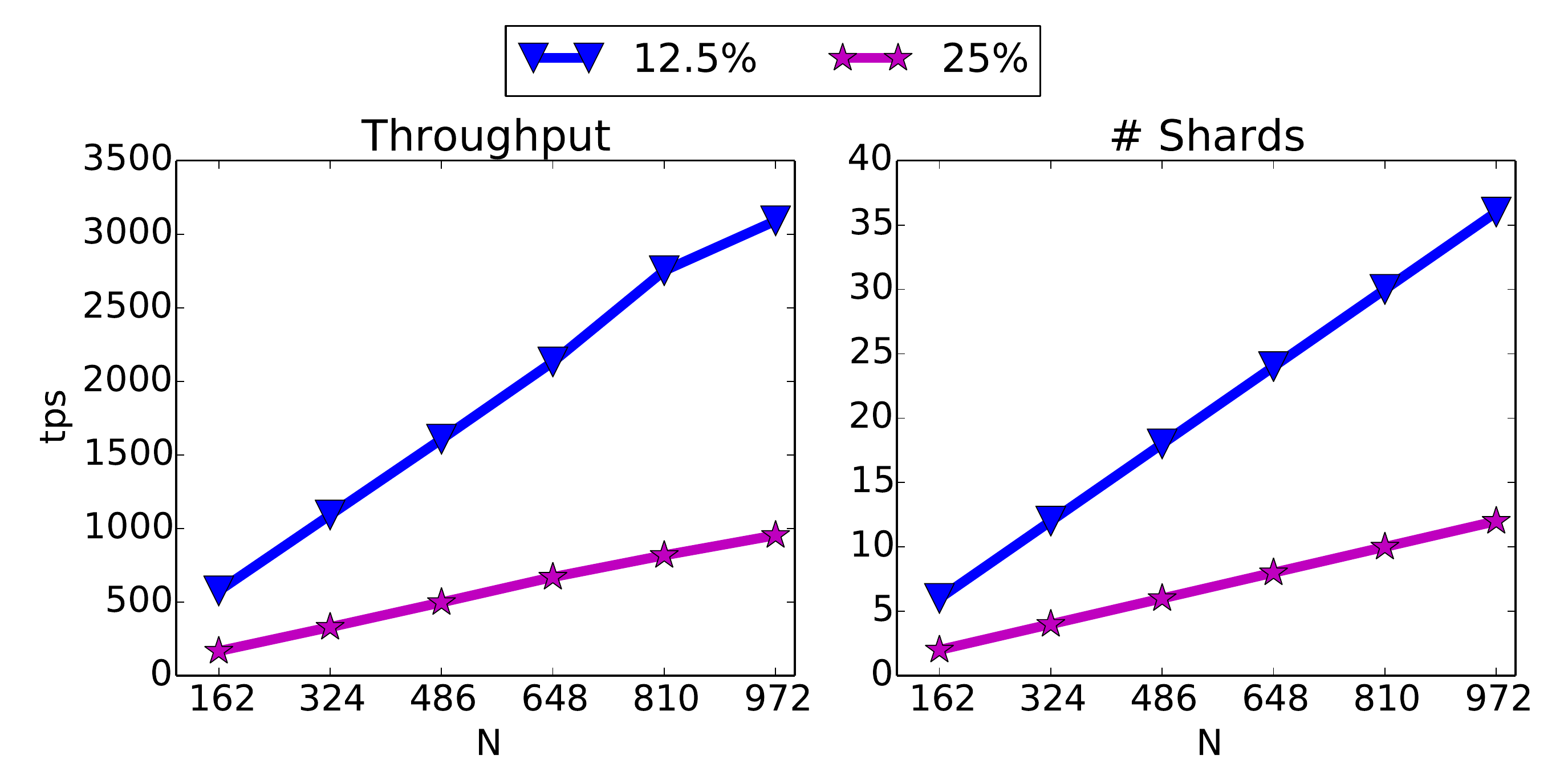}
\caption{{Sharding performance on GCP.}}
\label{fig:GCE_sharding}
\end{figure}

\subsection{Sharding performance}
\label{subsec:sharding_performance}
We first evaluate the performance of the coordination protocol by running many
shards on the local cluster with $f=1$. To saturate $S$ shards, we use $4S$
clients, each maintaining $128$ outstanding requests to the shard.
Figure~\ref{fig:cluster_sharding} (left) reports the throughput for Smallbank
with network size of up to $36$ nodes. The results for KVStore are similar and
are included in the Appendix. When HL-based sharding is used (as in OmniLedger),
there are up to 9 shards in total. Our sharding protocol uses AHL+, therefore
allowing for up to 12 shards. It can be seen that  the overall throughput of the
system scales linearly with the number of shards. Furthermore, when the
reference committee is involved, it becomes the bottleneck as the number of
shards grows. We vary the skewness factor (Zipf coefficient) of the workload,
and show the effect of contention on the overall throughput in
Figure~\ref{fig:cluster_sharding} (right).
As expected, the abort rate increases with the Zipf value.

Finally, we evaluate the throughput  scalability of our sharded blockchain system
 in a large-scale, realistic network environment. 
 For this experiment, we run Smallbank without the reference
committee, using up to $972$ nodes and $432$ clients spanning $8$ regions on
GCP. We consider two adversarial powers: $12.5\%$ and $25\%$ which are studied
in other sharded blockchains~\cite{omniledger, elastico}. 
The former requires a committee size of 27 nodes, and the
latter of 79 nodes to keep the probability of faulty committee below $2^{-20}$.
Figure~\ref{fig:GCE_sharding} shows the overall throughput and the corresponding
number of shards for the two adversarial powers. It can be seen that the
throughput scales linearly with the number of shards. In particular, for
$12.5\%$ adversary we achieve over $3,000$ transactions per second using $36$
shards, which is expected to grow higher with more shards\footnote{
We were unable to obtain enough resources to run with more shards at
 the time of writing.}.
For $25\%$ adversary, we observe a throughput of $954$ transactions per second
which is higher than that of OmniLedger with regular
validations~\cite{omniledger}.

\section{Related Works}
\label{sec:related_work}
We have discussed three related sharded blockchains, namely Elastico, OmniLedger and RapidChain, extensively
in the previous sections. Another related system is Chainspace~\cite{chainspace} which proposes a sharding
protocol for blockchain applications that are more general than cryptocurrencies. It allows smart contracts to
assign nodes to shards, as opposed to using a distributed shard formation protocol.  We do not consider
Chainspace in this work, due to its complex transaction model, complex coordination protocol, and low
throughput of only $300$ transactions per second even with $10$ shards.

\vspace{0.2cm}
\noindent{\bf Scaling blockchain with database techniques.} Various works have
exploited database techniques to improve aspects of the blockchain software
stack other than the consensus layer.
Forkbase~\cite{forkbase} is the first storage designed for blockchains,
supporting analytical queries at a much lower cost than the current key-value
backends. Dickerson {\em et al.}~\cite{dickerson17} add concurrency to Ethereum
execution engine by using  software transaction memory primitives. We expect
more works in improving the blockchain storage and execution engine. While
orthogonal to ours, they can be combined to enable scalable blockchains with
richer functionalities.

\vspace{0.2cm}
\noindent{\bf Off-chain scaling.} Instead of directly improving blockchain components, another approach
to scalability is to move as many transactions off the blockchain as possible.  Such {\em off-chain} solutions
allow users to execute transactions and reach consensus directly among each other, requiring minimal
interaction with the blockchain. The blockchain is used only for disputes and settlements.
Examples of off-chain solutions include payment channels~\cite{lightning} and state channels~\cite{sprites}.

\section{Conclusions}
\label{sec:conclusion}
In this paper, we applied database sharding techniques to blockchains. We
identified challenges that arise from the fundamental difference in failure
models between traditional distributed databases and blockchain systems. 
We addressed them by leveraging
TEEs to design fault-scalable consensus protocols and an efficient
shard formation protocol. Furthermore, we proposed a coordination protocol for cross-shard
transactions that supports general blockchain workloads. The coordination
protocol employs a Byzantine fault-tolerant reference committee to
guards against malicious coordinators.
Finally, we conducted extensive, large-scale evaluation of our designs in
realistic network settings, achieving over $3,000$ transactions per second with
many shards.

\bibliographystyle{ACM-Reference-Format}
\bibliography{paper}


\begin{thebibliography}{00}


\ifx \showCODEN    \undefined \def \showCODEN     #1{\unskip}     \fi
\ifx \showDOI      \undefined \def \showDOI       #1{#1}\fi
\ifx \showISBNx    \undefined \def \showISBNx     #1{\unskip}     \fi
\ifx \showISBNxiii \undefined \def \showISBNxiii  #1{\unskip}     \fi
\ifx \showISSN     \undefined \def \showISSN      #1{\unskip}     \fi
\ifx \showLCCN     \undefined \def \showLCCN      #1{\unskip}     \fi
\ifx \shownote     \undefined \def \shownote      #1{#1}          \fi
\ifx \showarticletitle \undefined \def \showarticletitle #1{#1}   \fi
\ifx \showURL      \undefined \def \showURL       {\relax}        \fi
\providecommand\bibfield[2]{#2}
\providecommand\bibinfo[2]{#2}
\providecommand\natexlab[1]{#1}
\providecommand\showeprint[2][]{arXiv:#2}

\bibitem[\protect\citeauthoryear{??}{sgx}{}]%
        {sgx_sdk}
\bibinfo{title}{Intel {SGX SDK} for {Linux}}.
\newblock \bibinfo{howpublished}{\url{https://github.com/01org/linux-sgx}}.
\newblock


\bibitem[\protect\citeauthoryear{??}{poe}{}]%
        {poet}
\bibinfo{title}{Proof of Elapsted Time}.
\newblock
  \bibinfo{howpublished}{\url{https://sawtooth.hyperledger.org/docs/core/releases/latest/introduction.html}}.
\newblock


\bibitem[\protect\citeauthoryear{??}{rip}{}]%
        {ripple}
\bibinfo{title}{{Ripple}}.
\newblock \bibinfo{howpublished}{\url{https://ripple.com}}.
\newblock


\bibitem[\protect\citeauthoryear{??}{msc}{}]%
        {mscoco}
\bibinfo{title}{{The Coco Framework}}.
\newblock \bibinfo{howpublished}{\url{http://aka.ms/cocopaper}}.
\newblock


\bibitem[\protect\citeauthoryear{??}{TPM}{}]%
        {TPM}
\bibinfo{title}{Trusted Computing Group}.
\newblock \bibinfo{howpublished}{\url{http://www.trustedcomputinggroup.org/}}.
\newblock


\bibitem[\protect\citeauthoryear{??}{sgx}{2018}]%
        {sgx_guide}
 \bibinfo{year}{2018}\natexlab{}.
\newblock \bibinfo{title}{Intel Software Guard Extensions Developer Guide}.
\newblock
  \bibinfo{howpublished}{\url{https://download.01.org/intel-sgx/linux-1.7/docs/Intel_SGX_Developer_Guide.pdf}}.
\newblock


\bibitem[\protect\citeauthoryear{Al-Bassam, Sonnino, Bano, Hrycyszyn, and
  Danezis}{Al-Bassam et~al\mbox{.}}{2017}]%
        {chainspace}
\bibfield{author}{\bibinfo{person}{Mustafa Al-Bassam}, \bibinfo{person}{Alberto
  Sonnino}, \bibinfo{person}{Shehar Bano}, \bibinfo{person}{Dave Hrycyszyn},
  {and} \bibinfo{person}{George Danezis}.} \bibinfo{year}{2017}\natexlab{}.
\newblock \showarticletitle{Chainspace: A Sharded Smart Contracts Platform}.
\newblock \bibinfo{journal}{{\em arXiv preprint arXiv:1708.03778\/}}
  (\bibinfo{year}{2017}).
\newblock


\bibitem[\protect\citeauthoryear{Alves and Felton}{Alves and Felton}{2004}]%
        {trustzone}
\bibfield{author}{\bibinfo{person}{Tiago Alves} {and} \bibinfo{person}{Don
  Felton}.} \bibinfo{year}{2004}\natexlab{}.
\newblock \bibinfo{booktitle}{{\em Trustzone: Integrated hardware and software
  security}}.
\newblock \bibinfo{type}{{T}echnical {R}eport}. \bibinfo{institution}{ARM}.
\newblock


\bibitem[\protect\citeauthoryear{Anati, Gueron, Johnson, and Scarlata}{Anati
  et~al\mbox{.}}{2013}]%
        {sgx_remote_attest}
\bibfield{author}{\bibinfo{person}{Ittai Anati}, \bibinfo{person}{Shay Gueron},
  \bibinfo{person}{Simon Johnson}, {and} \bibinfo{person}{Vincent Scarlata}.}
  \bibinfo{year}{2013}\natexlab{}.
\newblock \showarticletitle{Innovative technology for CPU based attestation and
  sealing}. In \bibinfo{booktitle}{{\em HASP}}.
\newblock


\bibitem[\protect\citeauthoryear{Behl, Distler, and Kapitza}{Behl
  et~al\mbox{.}}{2017}]%
        {hybster}
\bibfield{author}{\bibinfo{person}{Johannes Behl}, \bibinfo{person}{Tobias
  Distler}, {and} \bibinfo{person}{R{\"u}diger Kapitza}.}
  \bibinfo{year}{2017}\natexlab{}.
\newblock \showarticletitle{Hybrids on Steroids: SGX-Based High Performance
  BFT}. In \bibinfo{booktitle}{{\em EuroSys}}.
\newblock


\bibitem[\protect\citeauthoryear{{Bitcoin Wiki}}{{Bitcoin Wiki}}{2018}]%
        {bitcoinwiki_scalability}
\bibfield{author}{\bibinfo{person}{{Bitcoin Wiki}}.}
  \bibinfo{year}{2018}\natexlab{}.
\newblock \bibinfo{title}{Scalability}.
\newblock \bibinfo{howpublished}{en.bitcoin.it/wiki/scalability}.
\newblock


\bibitem[\protect\citeauthoryear{Brasser, Muller, Dmitrienko, Kostiainen,
  Capkun, and Sadeghi}{Brasser et~al\mbox{.}}{2017}]%
        {grandexp}
\bibfield{author}{\bibinfo{person}{Ferdinand Brasser}, \bibinfo{person}{Urs
  Muller}, \bibinfo{person}{Alexandra Dmitrienko}, \bibinfo{person}{Kari
  Kostiainen}, \bibinfo{person}{Srdjan Capkun}, {and}
  \bibinfo{person}{Ahmad-Reza Sadeghi}.} \bibinfo{year}{2017}\natexlab{}.
\newblock \showarticletitle{Software Grand Exposure: SGX Cache Attacks Are
  Practical}. In \bibinfo{booktitle}{{\em USENIX Security}}.
\newblock


\bibitem[\protect\citeauthoryear{Buterin}{Buterin}{2014}]%
        {eth_origin}
\bibfield{author}{\bibinfo{person}{Vitalik Buterin}.}
  \bibinfo{year}{2014}\natexlab{}.
\newblock \showarticletitle{Ethereum: A next-generation smart contract and
  decentralized application platform}.
\newblock \bibinfo{journal}{{\em
  \url{https://github.com/ethereum/wiki/wiki/White-Paper}\/}}
  (\bibinfo{year}{2014}).
\newblock


\bibitem[\protect\citeauthoryear{Castro}{Castro}{2001}]%
        {pbft_thesis}
\bibfield{author}{\bibinfo{person}{Miguel Castro}.}
  \bibinfo{year}{2001}\natexlab{}.
\newblock {\em \bibinfo{title}{Practical Byzantine Fault Tolerance}}.
\newblock \bibinfo{thesistype}{Ph.D. Dissertation}.
  \bibinfo{school}{Massachusetts Institute of Technolog}.
\newblock


\bibitem[\protect\citeauthoryear{Castro, Liskov, et~al\mbox{.}}{Castro
  et~al\mbox{.}}{1999}]%
        {pbft}
\bibfield{author}{\bibinfo{person}{Miguel Castro}, \bibinfo{person}{Barbara
  Liskov}, {et~al\mbox{.}}} \bibinfo{year}{1999}\natexlab{}.
\newblock \showarticletitle{Practical Byzantine fault tolerance}. In
  \bibinfo{booktitle}{{\em OSDI}}.
\newblock


\bibitem[\protect\citeauthoryear{Chen, Garfinkel, Lewis, Subrahmanyam,
  Waldspurger, Boneh, Dwoskin, and Ports}{Chen et~al\mbox{.}}{2008}]%
        {overshadow}
\bibfield{author}{\bibinfo{person}{Xiaoxin Chen}, \bibinfo{person}{Tal
  Garfinkel}, \bibinfo{person}{Christopher Lewis}, \bibinfo{person}{Pratap
  Subrahmanyam}, \bibinfo{person}{Carl~A. Waldspurger}, \bibinfo{person}{Dan
  Boneh}, \bibinfo{person}{Jeffrey Dwoskin}, {and} \bibinfo{person}{Dan~R.K.
  Ports}.} \bibinfo{year}{2008}\natexlab{}.
\newblock \showarticletitle{Overshadow: A Virtualization-Based Approach to
  Retrofitting Protection in Commodity Operating Systems}. In
  \bibinfo{booktitle}{{\em ASPLOS}}.
\newblock


\bibitem[\protect\citeauthoryear{Chun, Maniatis, Shenker, and Kubiatowicz}{Chun
  et~al\mbox{.}}{2007}]%
        {a2m}
\bibfield{author}{\bibinfo{person}{Byung-Gon Chun}, \bibinfo{person}{Petros
  Maniatis}, \bibinfo{person}{Scott Shenker}, {and} \bibinfo{person}{John
  Kubiatowicz}.} \bibinfo{year}{2007}\natexlab{}.
\newblock \showarticletitle{Attested append-only memory: Making adversaries
  stick to their word}. In \bibinfo{booktitle}{{\em OSR}}.
\newblock


\bibitem[\protect\citeauthoryear{Costan, Lebedev, and Devadas}{Costan
  et~al\mbox{.}}{}]%
        {sanctum}
\bibfield{author}{\bibinfo{person}{Victor Costan}, \bibinfo{person}{Ilia
  Lebedev}, {and} \bibinfo{person}{Srinivas Devadas}.}
\newblock \bibinfo{title}{Sanctum: Minimal hardware extensions for strong
  software isolation}.
\newblock \bibinfo{howpublished}{\url{https://eprint.iacr.org/2015/564.pdf}}.
\newblock


\bibitem[\protect\citeauthoryear{Dickerson, Gazzillo, Herlihy, and
  Koskinen}{Dickerson et~al\mbox{.}}{2017}]%
        {dickerson17}
\bibfield{author}{\bibinfo{person}{Thomas Dickerson}, \bibinfo{person}{Paul
  Gazzillo}, \bibinfo{person}{Maurice Herlihy}, {and} \bibinfo{person}{Eric
  Koskinen}.} \bibinfo{year}{2017}\natexlab{}.
\newblock \showarticletitle{Adding Concurrency to Smart Contracts}. In
  \bibinfo{booktitle}{{\em PODC}}.
\newblock


\bibitem[\protect\citeauthoryear{Dinh, Liu, Zhang, Chen, and Ooi}{Dinh
  et~al\mbox{.}}{2017a}]%
        {untangling_blockchain}
\bibfield{author}{\bibinfo{person}{Tien Tuan~Anh Dinh}, \bibinfo{person}{Rui
  Liu}, \bibinfo{person}{Meihui Zhang}, \bibinfo{person}{Gang Chen}, {and}
  \bibinfo{person}{Beng~Chin Ooi}.} \bibinfo{year}{2017}\natexlab{a}.
\newblock \showarticletitle{Untangling blockchain: a data processing view of
  blockchain systems}.
\newblock \bibinfo{journal}{{\em TKDE\/}} (\bibinfo{year}{2017}).
\newblock


\bibitem[\protect\citeauthoryear{Dinh, Wang, Chen, Liu, Ooi, and Tan}{Dinh
  et~al\mbox{.}}{2017b}]%
        {blockbench}
\bibfield{author}{\bibinfo{person}{Tien Tuan~Anh Dinh}, \bibinfo{person}{Ji
  Wang}, \bibinfo{person}{Gang Chen}, \bibinfo{person}{Rui Liu},
  \bibinfo{person}{Beng~Chin Ooi}, {and} \bibinfo{person}{Kian-Lee Tan}.}
  \bibinfo{year}{2017}\natexlab{b}.
\newblock \showarticletitle{BLOCKBENCH: A Framework for Analyzing Private
  Blockchains}. In \bibinfo{booktitle}{{\em SIGMOD}}.
\newblock


\bibitem[\protect\citeauthoryear{et~al.}{et~al.}{2012}]%
        {google_spanner}
\bibfield{author}{\bibinfo{person}{James C.~Corbett et al.}}
  \bibinfo{year}{2012}\natexlab{}.
\newblock \showarticletitle{Spanner: Google's Globally-Distributed Database}.
  In \bibinfo{booktitle}{{\em OSDI}}.
\newblock


\bibitem[\protect\citeauthoryear{{Fr8 Network}}{{Fr8 Network}}{2018}]%
        {fr8}
\bibfield{author}{\bibinfo{person}{{Fr8 Network}}.}
  \bibinfo{year}{2018}\natexlab{}.
\newblock \bibinfo{title}{Blockchain enabled logistic}.
\newblock \bibinfo{howpublished}{\url{https://fr8.network}}.
\newblock


\bibitem[\protect\citeauthoryear{Garcia-Molina}{Garcia-Molina}{2008}]%
        {2PC}
\bibfield{author}{\bibinfo{person}{Hector Garcia-Molina}.}
  \bibinfo{year}{2008}\natexlab{}.
\newblock \bibinfo{booktitle}{{\em Database systems: the complete book}}.
\newblock


\bibitem[\protect\citeauthoryear{Gervais, Karame, W{\"u}st, Glykantzis,
  Ritzdorf, and Capkun}{Gervais et~al\mbox{.}}{2016}]%
        {pow_sec_vs_tps}
\bibfield{author}{\bibinfo{person}{Arthur Gervais}, \bibinfo{person}{Ghassan~O
  Karame}, \bibinfo{person}{Karl W{\"u}st}, \bibinfo{person}{Vasileios
  Glykantzis}, \bibinfo{person}{Hubert Ritzdorf}, {and} \bibinfo{person}{Srdjan
  Capkun}.} \bibinfo{year}{2016}\natexlab{}.
\newblock \showarticletitle{On the security and performance of proof of work
  blockchains}. In \bibinfo{booktitle}{{\em CCS}}.
\newblock


\bibitem[\protect\citeauthoryear{Kogias, Jovanovic, Gailly, Khoffi, Gasser, and
  Ford}{Kogias et~al\mbox{.}}{2016}]%
        {byzcoin}
\bibfield{author}{\bibinfo{person}{Eleftherios~Kokoris Kogias},
  \bibinfo{person}{Philipp Jovanovic}, \bibinfo{person}{Nicolas Gailly},
  \bibinfo{person}{Ismail Khoffi}, \bibinfo{person}{Linus Gasser}, {and}
  \bibinfo{person}{Bryan Ford}.} \bibinfo{year}{2016}\natexlab{}.
\newblock \showarticletitle{Enhancing bitcoin security and performance with
  strong consistency via collective signing}. In \bibinfo{booktitle}{{\em
  USENIX Security}}.
\newblock


\bibitem[\protect\citeauthoryear{Kokoris-Kogias, Jovanovic, Gasser, Gailly, and
  Ford}{Kokoris-Kogias et~al\mbox{.}}{2017}]%
        {omniledger}
\bibfield{author}{\bibinfo{person}{Eleftherios Kokoris-Kogias},
  \bibinfo{person}{Philipp Jovanovic}, \bibinfo{person}{Linus Gasser},
  \bibinfo{person}{Nicolas Gailly}, {and} \bibinfo{person}{Bryan Ford}.}
  \bibinfo{year}{2017}\natexlab{}.
\newblock \showarticletitle{OmniLedger: A Secure, Scale-Out, Decentralized
  Ledger}.
\newblock \bibinfo{journal}{{\em IACR Cryptology ePrint Archive\/}}
  (\bibinfo{year}{2017}).
\newblock


\bibitem[\protect\citeauthoryear{Kwon}{Kwon}{}]%
        {tendermint}
\bibfield{author}{\bibinfo{person}{Jae Kwon}.}
\newblock \bibinfo{title}{Tendermint: Consensus without mining}.
\newblock
  \bibinfo{howpublished}{\url{https://tendermint.com/static/docs/tendermint.pdf}}.
\newblock


\bibitem[\protect\citeauthoryear{Lamport et~al\mbox{.}}{Lamport
  et~al\mbox{.}}{2001}]%
        {paxos}
\bibfield{author}{\bibinfo{person}{Leslie Lamport} {et~al\mbox{.}}}
  \bibinfo{year}{2001}\natexlab{}.
\newblock \showarticletitle{Paxos made simple}.
\newblock \bibinfo{journal}{{\em ACM Sigact News\/}} (\bibinfo{year}{2001}).
\newblock


\bibitem[\protect\citeauthoryear{Levin, Douceur, Lorch, and Moscibroda}{Levin
  et~al\mbox{.}}{2009}]%
        {trinc}
\bibfield{author}{\bibinfo{person}{Dave Levin}, \bibinfo{person}{John~R
  Douceur}, \bibinfo{person}{Jacob~R Lorch}, {and} \bibinfo{person}{Thomas
  Moscibroda}.} \bibinfo{year}{2009}\natexlab{}.
\newblock \showarticletitle{TrInc: Small Trusted Hardware for Large Distributed
  Systems}. In \bibinfo{booktitle}{{\em NSDI}}.
\newblock


\bibitem[\protect\citeauthoryear{{Lightning Network}}{{Lightning
  Network}}{2018}]%
        {lightning}
\bibfield{author}{\bibinfo{person}{{Lightning Network}}.}
  \bibinfo{year}{2018}\natexlab{}.
\newblock \bibinfo{title}{Lightning network: scalable, instant
  Bitcoin/blockchain transactions}.
\newblock \bibinfo{howpublished}{\url{https://lightning.network}}.
\newblock


\bibitem[\protect\citeauthoryear{Liu, Viotti, Cachin, Qu{\'e}ma, and
  Vukolic}{Liu et~al\mbox{.}}{2016}]%
        {xft}
\bibfield{author}{\bibinfo{person}{Shengyun Liu}, \bibinfo{person}{Paolo
  Viotti}, \bibinfo{person}{Christian Cachin}, \bibinfo{person}{Vivien
  Qu{\'e}ma}, {and} \bibinfo{person}{Marko Vukolic}.}
  \bibinfo{year}{2016}\natexlab{}.
\newblock \showarticletitle{XFT: Practical Fault Tolerance beyond Crashes}. In
  \bibinfo{booktitle}{{\em OSDI}}.
\newblock


\bibitem[\protect\citeauthoryear{Luu, Narayanan, Zheng, Baweja, Gilbert, and
  Saxena}{Luu et~al\mbox{.}}{2016}]%
        {elastico}
\bibfield{author}{\bibinfo{person}{Loi Luu}, \bibinfo{person}{Viswesh
  Narayanan}, \bibinfo{person}{Chaodong Zheng}, \bibinfo{person}{Kunal Baweja},
  \bibinfo{person}{Seth Gilbert}, {and} \bibinfo{person}{Prateek Saxena}.}
  \bibinfo{year}{2016}\natexlab{}.
\newblock \showarticletitle{A secure sharding protocol for open blockchains}.
  In \bibinfo{booktitle}{{\em CCS}}.
\newblock


\bibitem[\protect\citeauthoryear{Matetic, Ahmed, Kostiainen, Dhar, Sommer,
  Gervais, Juels, and Capkun}{Matetic et~al\mbox{.}}{2017}]%
        {matetic2017rote}
\bibfield{author}{\bibinfo{person}{Sinisa Matetic}, \bibinfo{person}{Mansoor
  Ahmed}, \bibinfo{person}{Kari Kostiainen}, \bibinfo{person}{Aritra Dhar},
  \bibinfo{person}{David Sommer}, \bibinfo{person}{Arthur Gervais},
  \bibinfo{person}{Ari Juels}, {and} \bibinfo{person}{Srdjan Capkun}.}
  \bibinfo{year}{2017}\natexlab{}.
\newblock \showarticletitle{ROTE: Rollback Protection for Trusted Execution}.
  In \bibinfo{booktitle}{{\em USENIX Security}}.
\newblock


\bibitem[\protect\citeauthoryear{McKeen, Alexandrovich, Berenzon, Rozas, Shafi,
  Shanbhogue, and Savagaonkar}{McKeen et~al\mbox{.}}{2013}]%
        {sgx}
\bibfield{author}{\bibinfo{person}{Frank McKeen}, \bibinfo{person}{Ilya
  Alexandrovich}, \bibinfo{person}{Alex Berenzon}, \bibinfo{person}{Carlos~V
  Rozas}, \bibinfo{person}{Hisham Shafi}, \bibinfo{person}{Vedvyas Shanbhogue},
  {and} \bibinfo{person}{Uday~R Savagaonkar}.} \bibinfo{year}{2013}\natexlab{}.
\newblock \showarticletitle{Innovative instructions and software model for
  isolated execution}. In \bibinfo{booktitle}{{\em HASP}}.
\newblock


\bibitem[\protect\citeauthoryear{Medilot}{Medilot}{2018}]%
        {medilot}
\bibfield{author}{\bibinfo{person}{Medilot}.} \bibinfo{year}{2018}\natexlab{}.
\newblock \bibinfo{title}{Transforming healthcare for all}.
\newblock \bibinfo{howpublished}{\url{https://medilot.com}}.
\newblock


\bibitem[\protect\citeauthoryear{Micali}{Micali}{2016}]%
        {algorand}
\bibfield{author}{\bibinfo{person}{Silvio Micali}.}
  \bibinfo{year}{2016}\natexlab{}.
\newblock \showarticletitle{ALGORAND: the efficient and democratic ledger}.
\newblock \bibinfo{journal}{{\em arXiv preprint arXiv:1607.01341\/}}
  (\bibinfo{year}{2016}).
\newblock


\bibitem[\protect\citeauthoryear{Micali, Rabin, and Vadhan}{Micali
  et~al\mbox{.}}{1999}]%
        {vrf}
\bibfield{author}{\bibinfo{person}{Silvio Micali}, \bibinfo{person}{Michael
  Rabin}, {and} \bibinfo{person}{Salil Vadhan}.}
  \bibinfo{year}{1999}\natexlab{}.
\newblock \showarticletitle{Verifiable random functions}. In
  \bibinfo{booktitle}{{\em FOCS}}.
\newblock


\bibitem[\protect\citeauthoryear{Miller, Bentov, Kumaresan, Cordi, and
  McCorry}{Miller et~al\mbox{.}}{2017}]%
        {sprites}
\bibfield{author}{\bibinfo{person}{Andrew Miller}, \bibinfo{person}{Iddo
  Bentov}, \bibinfo{person}{Ranjit Kumaresan}, \bibinfo{person}{Christopher
  Cordi}, {and} \bibinfo{person}{Patrick McCorry}.}
  \bibinfo{year}{2017}\natexlab{}.
\newblock \bibinfo{title}{Sprites and state channels: payment networks that go
  faster than lightning}.
\newblock \bibinfo{howpublished}{\url{https://arxiv.org/pdf/1702.05812.pdf}}.
\newblock


\bibitem[\protect\citeauthoryear{Mu, Nelson, Lloyd, and Li}{Mu
  et~al\mbox{.}}{2016}]%
        {janus}
\bibfield{author}{\bibinfo{person}{Shuai Mu}, \bibinfo{person}{Lamont Nelson},
  \bibinfo{person}{Wyatt Lloyd}, {and} \bibinfo{person}{Jinyang Li}.}
  \bibinfo{year}{2016}\natexlab{}.
\newblock \showarticletitle{Consolidating Concurrency Control and Consensus for
  Commits under Conflicts}. In \bibinfo{booktitle}{{\em OSDI}}.
\newblock


\bibitem[\protect\citeauthoryear{Nakamoto}{Nakamoto}{2008}]%
        {btc_origin}
\bibfield{author}{\bibinfo{person}{Satoshi Nakamoto}.}
  \bibinfo{year}{2008}\natexlab{}.
\newblock \bibinfo{title}{Bitcoin: A peer-to-peer electronic cash system}.
\newblock
\newblock


\bibitem[\protect\citeauthoryear{Ongaro and Ousterhout}{Ongaro and
  Ousterhout}{2014}]%
        {raft}
\bibfield{author}{\bibinfo{person}{Diego Ongaro} {and} \bibinfo{person}{John~K
  Ousterhout}.} \bibinfo{year}{2014}\natexlab{}.
\newblock \showarticletitle{In search of an understandable consensus
  algorithm}. In \bibinfo{booktitle}{{\em USENIX ATC}}.
\newblock


\bibitem[\protect\citeauthoryear{Pass, Seeman, and Shelat}{Pass
  et~al\mbox{.}}{2017}]%
        {bitcoin_async}
\bibfield{author}{\bibinfo{person}{Rafael Pass}, \bibinfo{person}{Lior Seeman},
  {and} \bibinfo{person}{Abhi Shelat}.} \bibinfo{year}{2017}\natexlab{}.
\newblock \showarticletitle{Analysis of the blockchain protocol in asynchronous
  networks}. In \bibinfo{booktitle}{{\em EUROCRYPT}}.
\newblock


\bibitem[\protect\citeauthoryear{Syta, Jovanovic, Kogias, Gailly, Gasser,
  Khoffi, Fischer, and Ford}{Syta et~al\mbox{.}}{2017}]%
        {randhound}
\bibfield{author}{\bibinfo{person}{Ewa Syta}, \bibinfo{person}{Philipp
  Jovanovic}, \bibinfo{person}{Eleftherios~Kokoris Kogias},
  \bibinfo{person}{Nicolas Gailly}, \bibinfo{person}{Linus Gasser},
  \bibinfo{person}{Ismail Khoffi}, \bibinfo{person}{Michael~J Fischer}, {and}
  \bibinfo{person}{Bryan Ford}.} \bibinfo{year}{2017}\natexlab{}.
\newblock \showarticletitle{Scalable bias-resistant distributed randomness}. In
  \bibinfo{booktitle}{{\em IEEE S\& P}}.
\newblock


\bibitem[\protect\citeauthoryear{Tramer, Zhang, Lin, Hubaux, Juels, and
  Shi}{Tramer et~al\mbox{.}}{2017}]%
        {sealed_glass_proof}
\bibfield{author}{\bibinfo{person}{Florian Tramer}, \bibinfo{person}{Fan
  Zhang}, \bibinfo{person}{Huang Lin}, \bibinfo{person}{Jean-Pierre Hubaux},
  \bibinfo{person}{Ari Juels}, {and} \bibinfo{person}{Elaine Shi}.}
  \bibinfo{year}{2017}\natexlab{}.
\newblock \showarticletitle{Sealed-glass proofs: Using transparent enclaves to
  prove and sell knowledge}. In \bibinfo{booktitle}{{\em EuroS\&P}}.
\newblock


\bibitem[\protect\citeauthoryear{Van~Bulck, Minkin, Weisse, Genkin, Kasikci,
  Piessens, Silberstein, Wenisch, Yarom, and Strackx}{Van~Bulck
  et~al\mbox{.}}{2018}]%
        {foreshadow}
\bibfield{author}{\bibinfo{person}{Jo Van~Bulck}, \bibinfo{person}{Marina
  Minkin}, \bibinfo{person}{Ofir Weisse}, \bibinfo{person}{Daniel Genkin},
  \bibinfo{person}{Baris Kasikci}, \bibinfo{person}{Frank Piessens},
  \bibinfo{person}{Mark Silberstein}, \bibinfo{person}{Thomas~F Wenisch},
  \bibinfo{person}{Yuval Yarom}, {and} \bibinfo{person}{Raoul Strackx}.}
  \bibinfo{year}{2018}\natexlab{}.
\newblock \showarticletitle{FORESHADOW: Extracting the Keys to the Intel SGX
  Kingdom with Transient Out-of-Order Execution}. In \bibinfo{booktitle}{{\em
  USENIX Security}}.
\newblock


\bibitem[\protect\citeauthoryear{Wang, Dinh, Lin, Xie, Zhang, Cai, Chen, Fu,
  Ooi, and Ruan}{Wang et~al\mbox{.}}{2018}]%
        {forkbase}
\bibfield{author}{\bibinfo{person}{Sheng Wang}, \bibinfo{person}{Tien Tuan~Anh
  Dinh}, \bibinfo{person}{Qian Lin}, \bibinfo{person}{Zhongle Xie},
  \bibinfo{person}{Meihui Zhang}, \bibinfo{person}{Qingchao Cai},
  \bibinfo{person}{Gang Chen}, \bibinfo{person}{Wanzeng Fu},
  \bibinfo{person}{Beng~Chin Ooi}, {and} \bibinfo{person}{Pingcheng Ruan}.}
  \bibinfo{year}{2018}\natexlab{}.
\newblock \showarticletitle{ForkBase: An Efficient Storage Engine for
  Blockchain and Forkable Applications}. In \bibinfo{booktitle}{{\em VLDB}}.
\newblock


\bibitem[\protect\citeauthoryear{Yan, Yang, Zhang, Lin, Wong, Salem, and
  Brecht}{Yan et~al\mbox{.}}{2018}]%
        {carousel}
\bibfield{author}{\bibinfo{person}{Xinan Yan}, \bibinfo{person}{Linguan Yang},
  \bibinfo{person}{Hongbo Zhang}, \bibinfo{person}{Xiayue~Charles Lin},
  \bibinfo{person}{Bernard Wong}, \bibinfo{person}{Kenneth Salem}, {and}
  \bibinfo{person}{Tim Brecht}.} \bibinfo{year}{2018}\natexlab{}.
\newblock \showarticletitle{Carousel: low-latency transaction processing for
  globally-distributed data}. In \bibinfo{booktitle}{{\em SIGMOD}}.
\newblock


\bibitem[\protect\citeauthoryear{Zamani, Movahedi, and Raykova}{Zamani
  et~al\mbox{.}}{2018}]%
        {rapidchain}
\bibfield{author}{\bibinfo{person}{Mahdi Zamani}, \bibinfo{person}{Mahnush
  Movahedi}, {and} \bibinfo{person}{Mariana Raykova}.}
  \bibinfo{year}{2018}\natexlab{}.
\newblock \showarticletitle{RapidChain: Scaling Blockchain via Full Sharding}.
  In \bibinfo{booktitle}{{\em CCS}}.
\newblock


\bibitem[\protect\citeauthoryear{Zhang, Sharma, Szekeres, Krishnamurthy, and
  Ports}{Zhang et~al\mbox{.}}{2015}]%
        {tapir}
\bibfield{author}{\bibinfo{person}{Irene Zhang}, \bibinfo{person}{Naveen~Kr.
  Sharma}, \bibinfo{person}{Adriana Szekeres}, \bibinfo{person}{Arvind
  Krishnamurthy}, {and} \bibinfo{person}{Dan R.~K. Ports}.}
  \bibinfo{year}{2015}\natexlab{}.
\newblock \showarticletitle{Building Consistent Transactions with Inconsistent
  Replication}. In \bibinfo{booktitle}{{\em SOSP}}.
\newblock


\end{thebibliography}

\appendix
\section{Defenses against Rollback Attacks}
\label{apd:rollback_attacks}
Data sealing mechanism enables enclaves to save their states to persistent
storage, allowing them to resume their operations upon recovery.
However, enclave recovery is vulnerable rollback attack~\cite{matetic2017rote}.

\vspace{2mm}
\noindent \textbf{AHL+}. The adversary can cause the enclave of  AHL+ to
restart, and supply it with a \textit{stale} log heads upon its resumption. The
enclave resuming with stale log heads ``forgets'' all  messages appended after the stale
log heads, allowing the adversary to equivocate.

Denote by $H$ the sequence number of the last consensus message the enclave
processes prior to its restart. The recovering enclave must not accept any
message with a sequence number lower than or equal to $H$.  We derive an
estimation procedure that allows the resuming enclave to estimate an upper
bound, $H_M$, on the latest sequence number it would have observed if it were
not crashed. The goal of this estimation is to guarantee that $H_M \geq H$,
ensuring protocol's safety.

The enclave starts the estimation procedure by querying all its peers for the
sequence number of their last checkpoint, denoted by $ckp$.
The resuming enclave uses the responses to select $ckp_M$, which is a value
$ckp$ it receives from one node $j$ such that there are $f$ replicas other than
$j$ reporting values less than or equal to $ckp_M$. It then sets the value $H_M$
to $H_M = L + ckp_M$ where $L$ is a preset difference between the node's high
and low watermarks. The test against $ckp$ responses of $f$ other replicas
ensures that $ckp_M$ is greater than the sequence number of any stable
checkpoint the resuming enclave may have; otherwise, there must be at least $f$
$ckp$ responses that are larger than $ckp_M$, which is not possible due to
quorum intersection.

The resuming enclave will not append any message to its logs until it is {\it
fully recovered}. This effectively refrains its host node from sending any
message or processing any request, for the node cannot obtain the proof of
append operation generated by the enclave. The enclave is fully recovered only
after it is presented with a correct stable checkpoint with a sequence number
greater than or equal to $H_M$. At this point, it starts accepting append
operations, and the host node can actively participate in the protocol. Since
$H_M$ is an upper bound on the sequence number the AHL+ enclave would observe
had it not been crashed, and that the host node cannot send any message with a
sequence number lower than $H_M$ once its enclave is restarted, the protocol is
safe from equivocation.


\vspace{2mm}
\noindent \textbf{\textproc{RandomnessBeacon}}. The random values \texttt{q} and
\texttt{rnd} are bound to the epoch number $e$ and a counter $v$ to prevent the
adversary from selectively discarding the enclave's output to bias the
randomness. These values, nonetheless, are stored in the enclave's volatile
memory. The adversary may attempt to restart the enclave and invoke it using the
same epoch number $e$ to collect different values of \texttt{q} and
\texttt{rnd}. Fortunately, the adversary only has a window of $\Delta$ from the
beginning of epoch $e$ to bias its \texttt{q} and \texttt{rnd} in that same
epoch (after $\Delta$, nodes have already locked the value of \texttt{rnd} used
in  epoch $e$). Thus, to prevent the adversary from restarting the enclave to
bias \texttt{q} and \texttt{rnd}, it suffices to bar the enclave from issuing
these two random values for any input $e \neq 0$  for a duration of $\Delta$
since its instantiation. The genesis epoch requires a more subtle set-up wherein
participants are forced to not restart their enclaves during that first epoch.
This can be realized by involving the use of CPU's monotonic-counter. Such
process needs to be conducted only once at the system's bootstrap.

%
\section{Probability of Cross-Shard Transactions}
\label{apd:cross_shard_prob}
We examine the probability that a transaction is cross-shard (i.e., it affects
multiple shards' states at the same time). Consider a $d$-argument transaction
$tx$ that affects the values (states) of $d$ different arguments.
Without loss of generality, let us assume that arguments are mapped to shards
uniformly at random, based on the randomness provided by a cryptographic hash
function applied on the arguments. Let $k$ be the total number of shards formed
in the system. The probability that the transaction $tx$ affects the states of
exactly $x \leq min(d, k)$ shards can be calculated based on the multinomial
distribution as follows:

\begin{equation}	
\prod_{i=1}^{x-1} \frac{k-i}{k} \sum_{p_1 + p_2 + p_x = d-x} \prod_{j=1}^{x} (\frac{j}{k})^{p_j}
\end{equation}	

While OmniLedger and RapidChain give a similar calculation, they only consider a
specific type of UTXO transactions whose outputs are all managed by a single
output committee. Unfortunately, such calculation does not extend to UTXO
transactions whose outputs belong to separate committees, let alone non-UTXO
distributed transactions.



\begin{table*} \centering
\caption{Latency (ms) between different regions on Google Cloud Platform.}
\label{table:lat-gce}
\resizebox{0.9\textwidth}{!} {
\begin{tabular}{|l|r|r|r|r|r|r|r|r|}
\hline
\textbf{Zone} & \textbf{us-west1-b} & \textbf{us-west2-a} & \textbf{us-east1-b} & \textbf{us-east4-b} & \textbf{asia-east1-b} & \textbf{asia-southeast1-b} & \textbf{europe-west1-b} & \textbf{europe-west2-a} \\
\hline\hline
\textbf{us-west1-b} & 0.0 & \cellcolor{grad_4} 24.7 & \cellcolor{grad_3} 66.7 & \cellcolor{grad_4} 59.0 & \cellcolor{grad_2} 120.2 & \cellcolor{grad_2} 150.8 & \cellcolor{grad_2} 138.9 & \cellcolor{grad_2} 132.7 \\
\hline
\textbf{us-west2-a} & \cellcolor{grad_4} 24.7 & 0.0 & \cellcolor{grad_4} 62.9 & \cellcolor{grad_4} 60.5 & \cellcolor{grad_2} 129.5 & \cellcolor{grad_2} 160.5 & \cellcolor{grad_2} 140.4 & \cellcolor{grad_2} 136.1 \\
\hline
\textbf{us-east1-b} & \cellcolor{grad_3} 66.7 & \cellcolor{grad_4} 62.9 & 0.0 & \cellcolor{grad_4} 12.7 & \cellcolor{grad_1} 183.8 & \cellcolor{grad_1} 216.6 & \cellcolor{grad_3} 93.1 & \cellcolor{grad_3} 88.2 \\
\hline
\textbf{us-east4-b} & \cellcolor{grad_4} 59.1 & \cellcolor{grad_4} 60.4 & \cellcolor{grad_4} 12.7 & 0.0 & \cellcolor{grad_1} 176.6 & \cellcolor{grad_1} 208.4 & \cellcolor{grad_3} 81.9 & \cellcolor{grad_3} 75.6 \\
\hline
\textbf{asia-east1-b} & \cellcolor{grad_3} 118.7 & \cellcolor{grad_2} 129.5 & \cellcolor{grad_1} 184.9 & \cellcolor{grad_1} 176.6 & 0.0 & \cellcolor{grad_4} 50.5 & \cellcolor{grad_0} 255.5 & \cellcolor{grad_0} 252.5 \\
\hline
\textbf{asia-southeast1-b} & \cellcolor{grad_2} 150.8 & \cellcolor{grad_2} 160.5 & \cellcolor{grad_1} 216.7 & \cellcolor{grad_1} 208.3 & \cellcolor{grad_4} 50.6 & 0.0 & \cellcolor{grad_0} 288.8 & \cellcolor{grad_0} 283.8 \\
\hline
\textbf{europe-west1-b} & \cellcolor{grad_2} 138.9 & \cellcolor{grad_2} 140.5 & \cellcolor{grad_3} 93.2 & \cellcolor{grad_3} 81.8 & \cellcolor{grad_0} 255.7 & \cellcolor{grad_0} 288.7 & 0.0 & \cellcolor{grad_4} 7.1 \\
\hline
\textbf{europe-west2-a} & \cellcolor{grad_2} 132.1 & \cellcolor{grad_2} 134.9 & \cellcolor{grad_3} 88.1 & \cellcolor{grad_3} 76.6 & \cellcolor{grad_0} 252.1 & \cellcolor{grad_0} 283.9 & \cellcolor{grad_4} 7.1 & 0.0 \\
\hline
\end{tabular}
}
\end{table*}

\section{Additional Evaluation Results}
\label{sec:apd_experiments}
\begin{figure}
\centering
\includegraphics[width=0.45\textwidth]{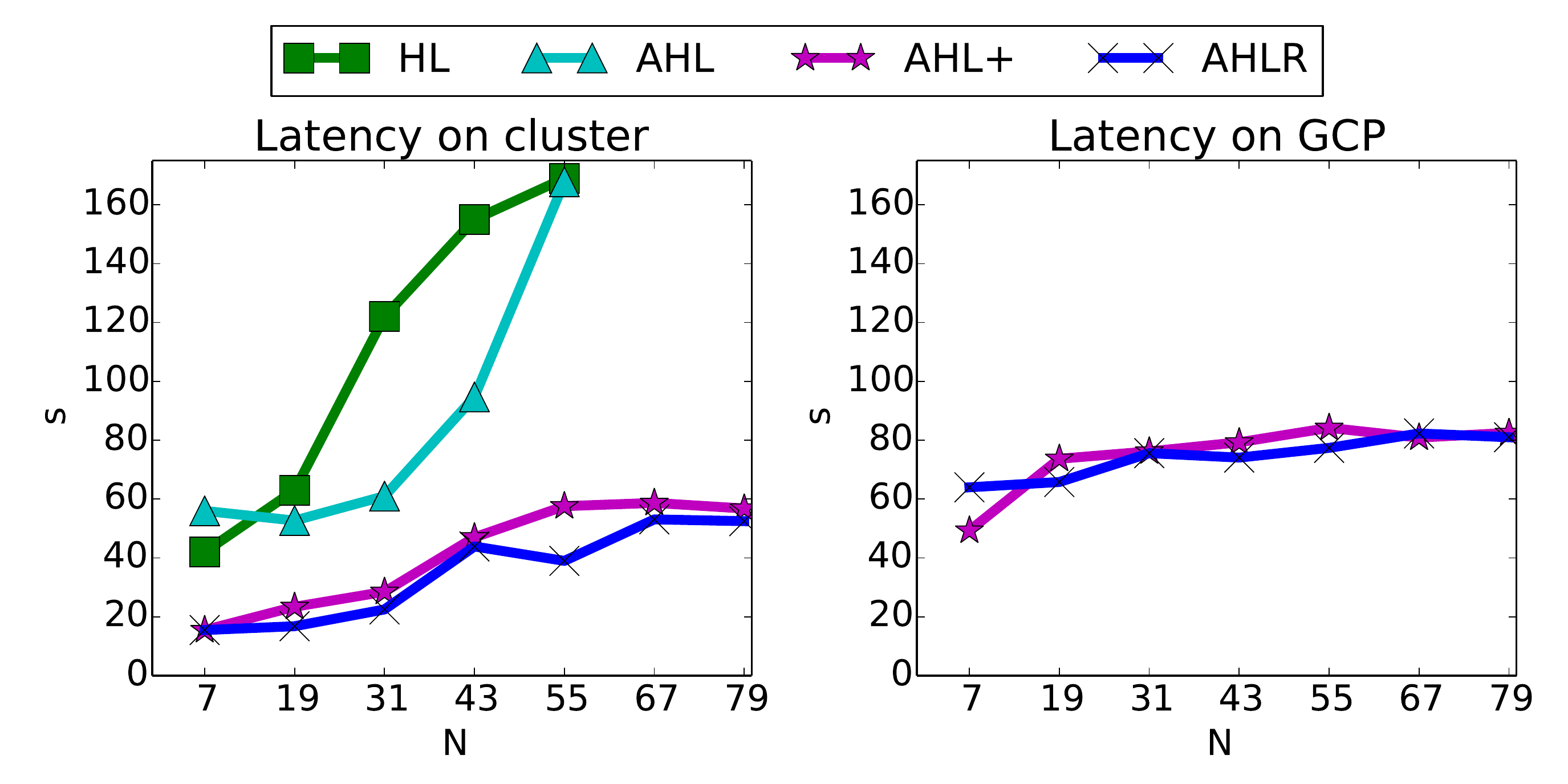}
\caption{AHL+ latency on local cluster and GCP.}
\label{fig:pbft_latency}
\end{figure}

\begin{figure}
\centering
\includegraphics[width=0.45\textwidth]{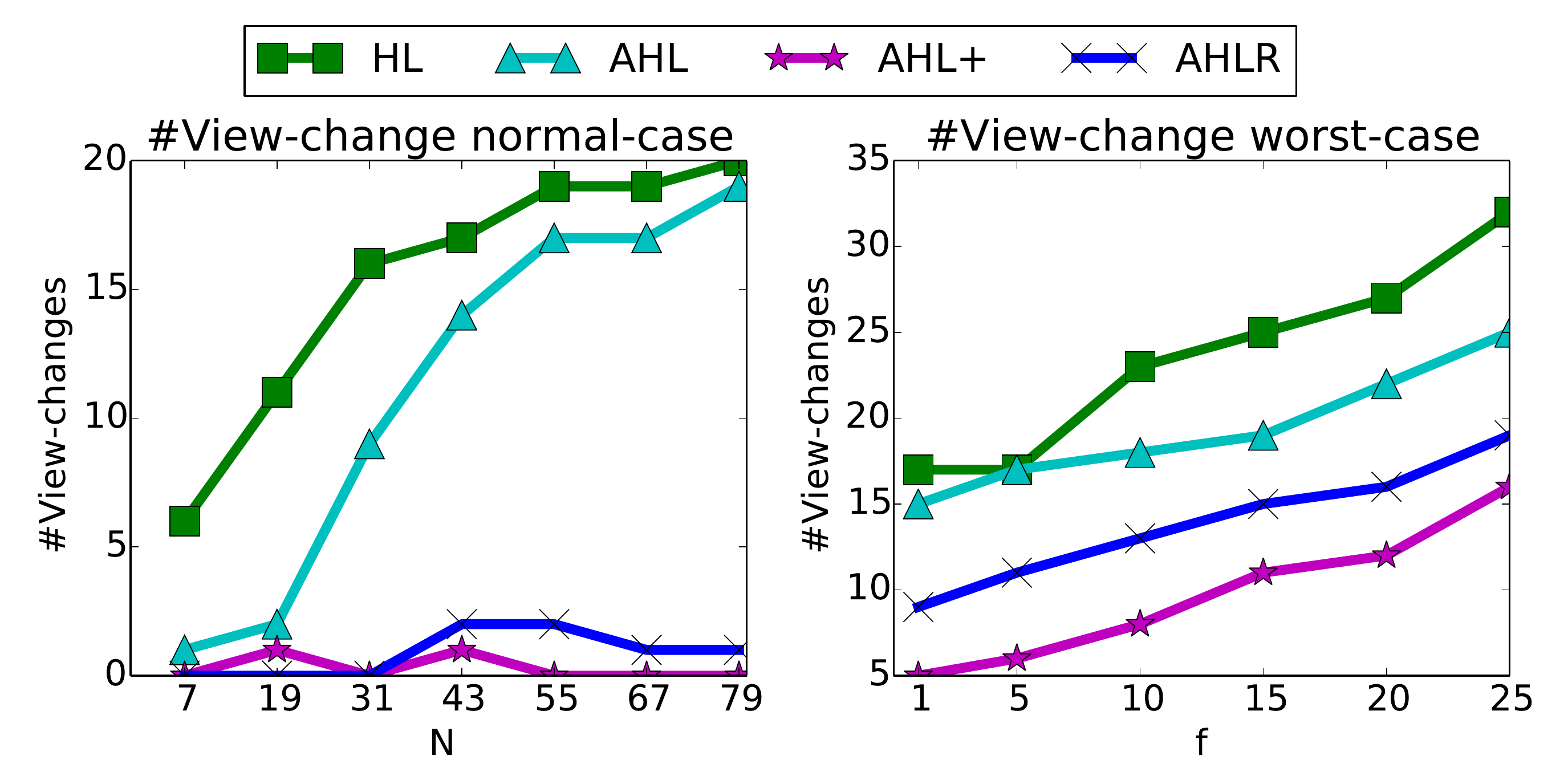}
\caption{\# View-changes of AHL+ on local cluster.}
\label{fig:pbft_vc}
\end{figure}

\begin{figure}
\centering
\includegraphics[width=0.45\textwidth]{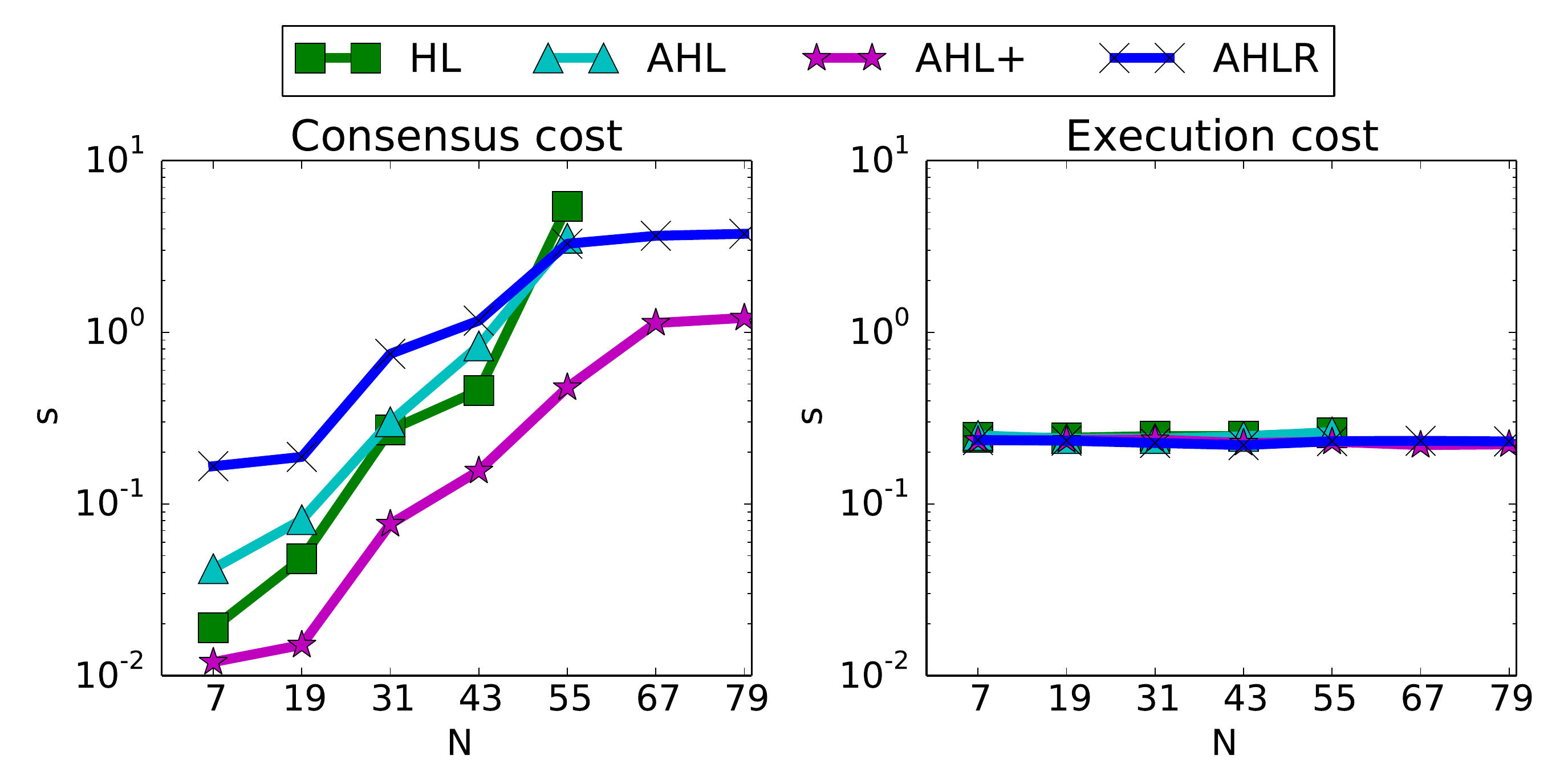}
\caption{Consensus and execution cost breakdown of AHL+ on local cluster.}
\label{fig:consensus_cost_breakdown}
\end{figure}

\begin{figure}
\centering
\includegraphics[width=0.235\textwidth]{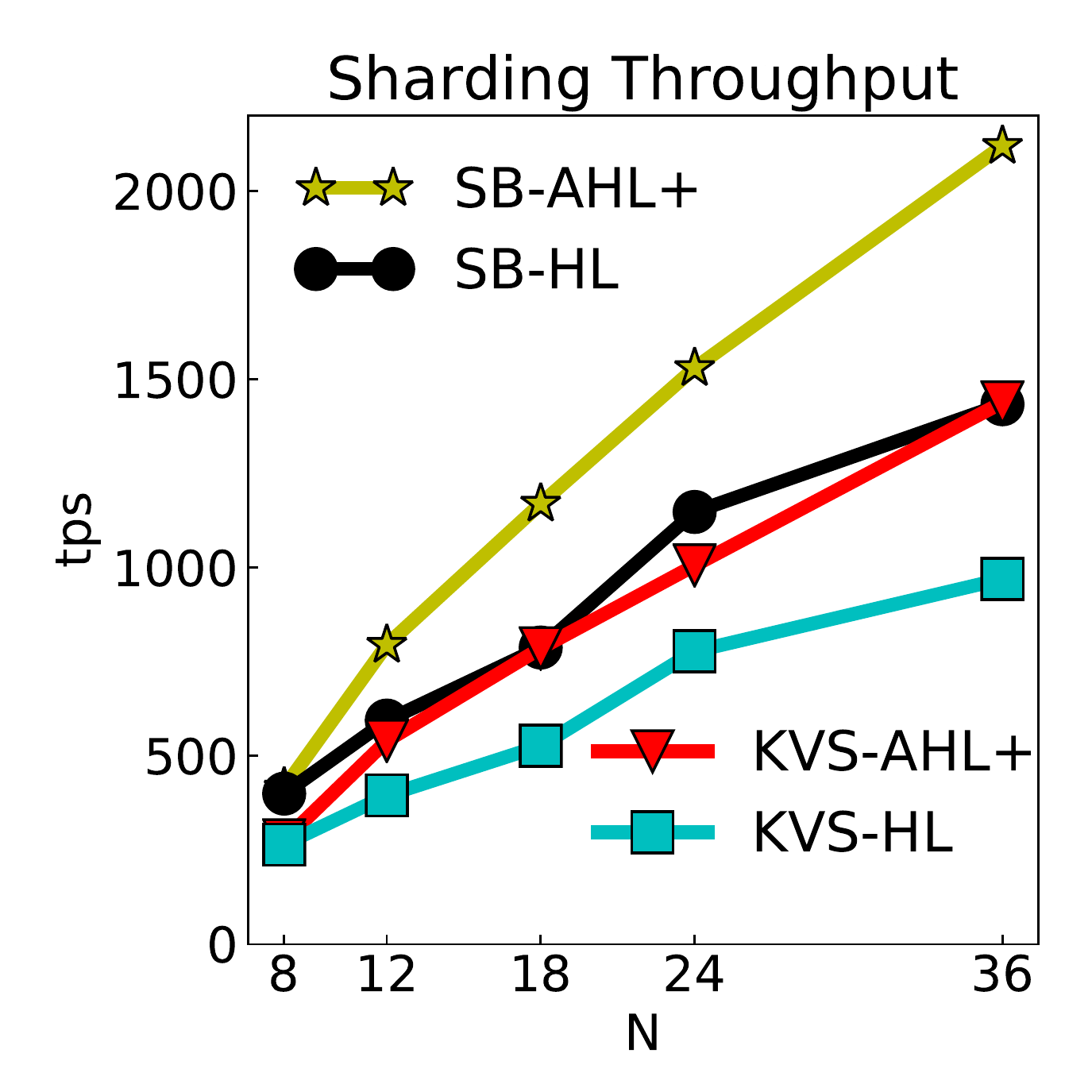}
\caption{Sharding with KVStore vs. Smallbank.}
\label{fig:kv_store_sharding}
\end{figure}

\begin{figure}
\centering
\includegraphics[width=0.45\textwidth]{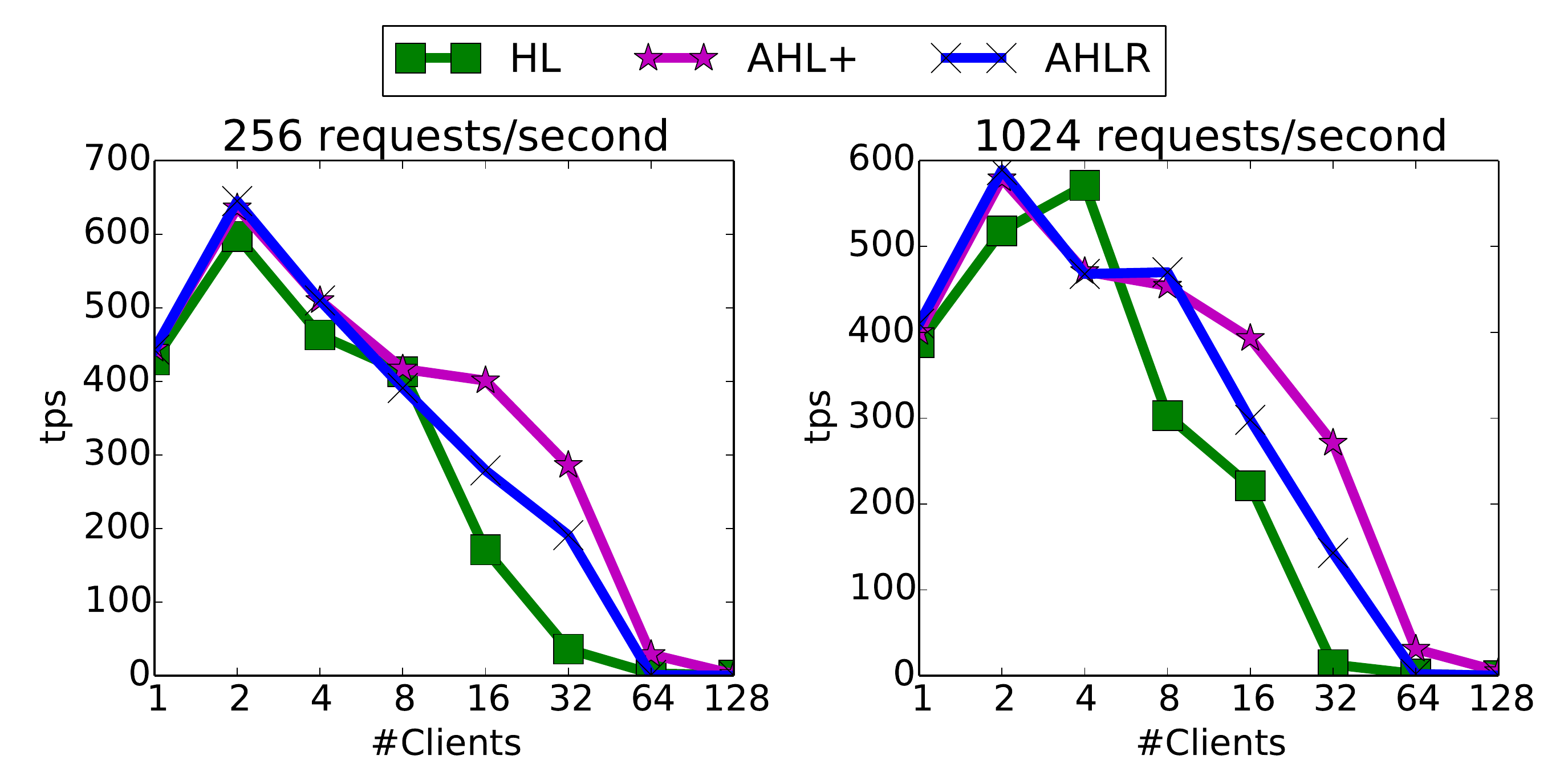}
\caption{Throughput with varying workload on GCP.}
\label{fig:vary_workload_GCP}
\end{figure}

\begin{figure}
\centering
\includegraphics[width=0.4\textwidth]{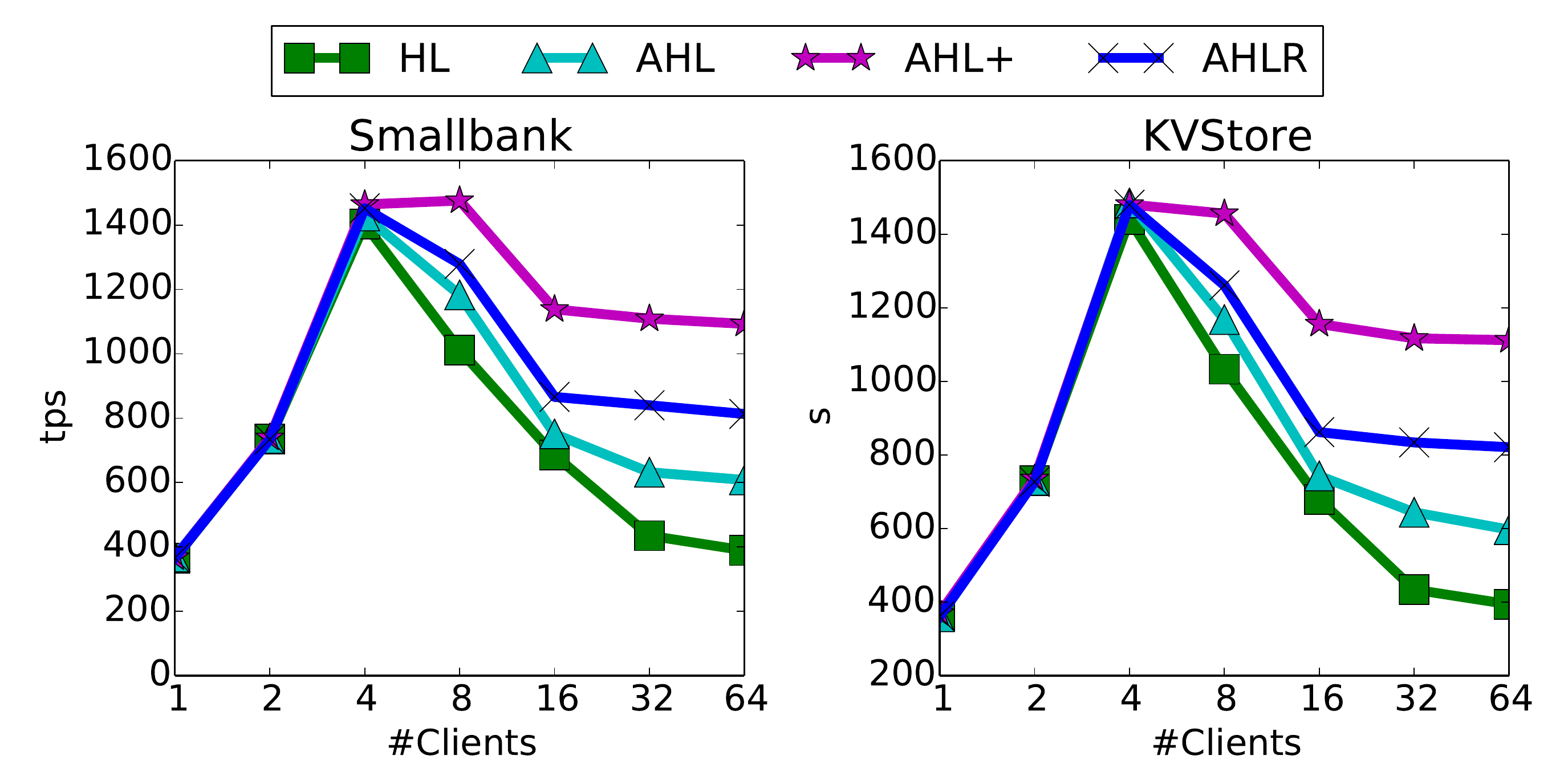}
\caption{Throughput with varying workload on local cluster.}
\label{fig:vary_workload_cluster}
\end{figure}

This section provides additional results to those discussed in
Section~\ref{sec:eval}. First, the latency among the 8 GPC regions used in our
experiments is listed in Table~\ref{table:lat-gce}.
Figure~\ref{fig:pbft_latency} and \ref{fig:pbft_vc} show the latency and number
of view-changes in different consensus protocols as the network size increases.
Figure~\ref{fig:consensus_cost_breakdown} presents the cost breakdown for a
block of transactions, showing that the cost of transaction execution is an
order of magnitude smaller than that of consensus.
Figure~\ref{fig:vary_workload_GCP} and \ref{fig:vary_workload_cluster}
demonstrate   AHL+'s throughput with varying numbers of clients on the cluster
and on GCP.
Figure~\ref{fig:kv_store_sharding} compares the sharding throughput under
KVStore versus Smallbank.

\subsection{\spoet\ vs. \poet}
We evaluate the performance of \poet\ and \spoet based on Hyperledger Sawtooth
v0.8 implementation. On the local cluster, we run 4 nodes on each physical
server, and impose 50 Mbps bandwidth limit and 100ms latency on the network
links. On GCP, we run each node on an instance with 2 vCPUs, and the nodes are
distributed over $8$ regions. We set $l = \frac{\log(N)}{2}$, reducing the
effective network size to $\sqrt{N}$.

\begin{figure}
\includegraphics[width=0.45\textwidth]{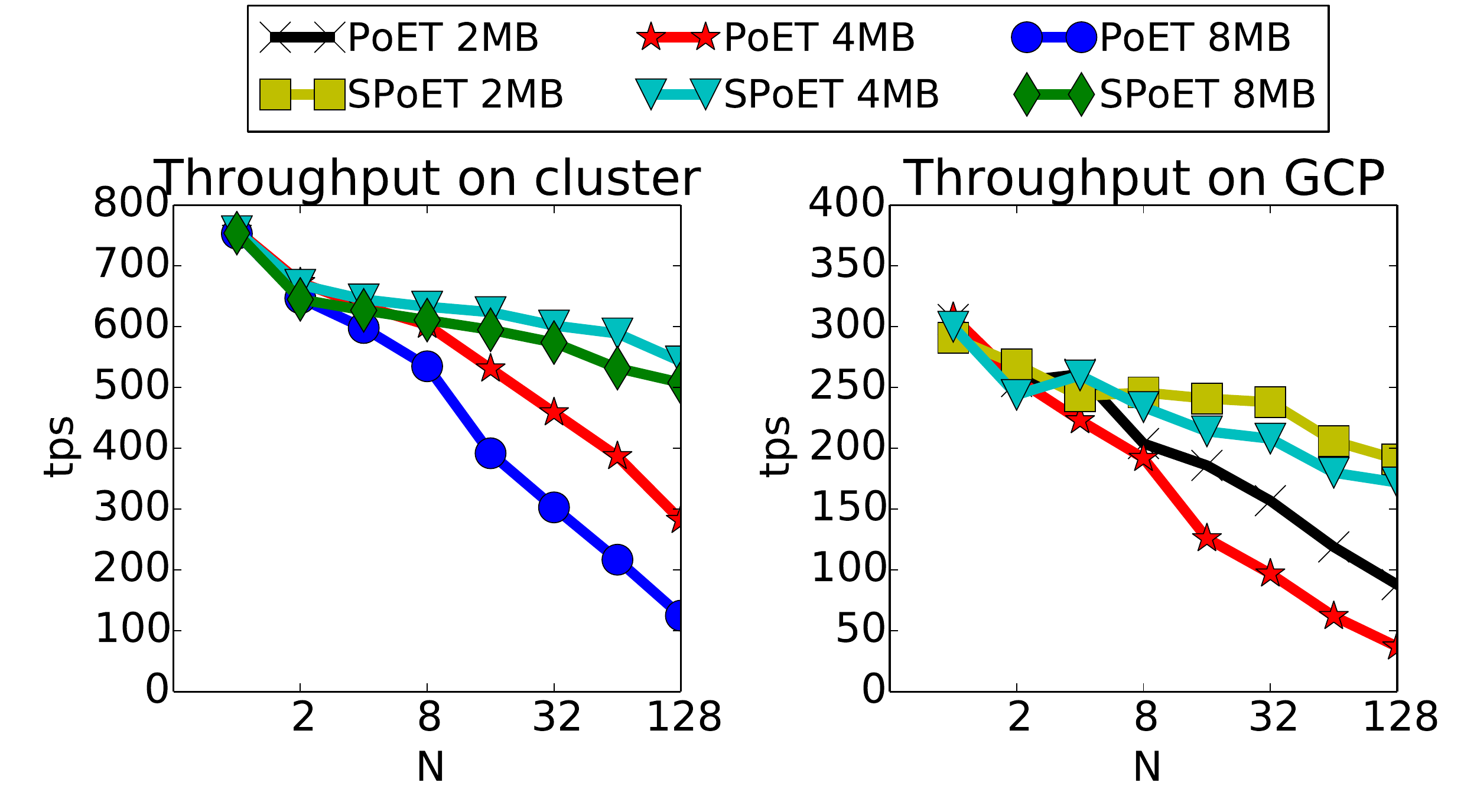}
\caption{\poet\ and \spoet\ performance.}
\label{fig:spoet}
\end{figure}

\begin{figure}
\centering
\includegraphics[width=0.45\textwidth]{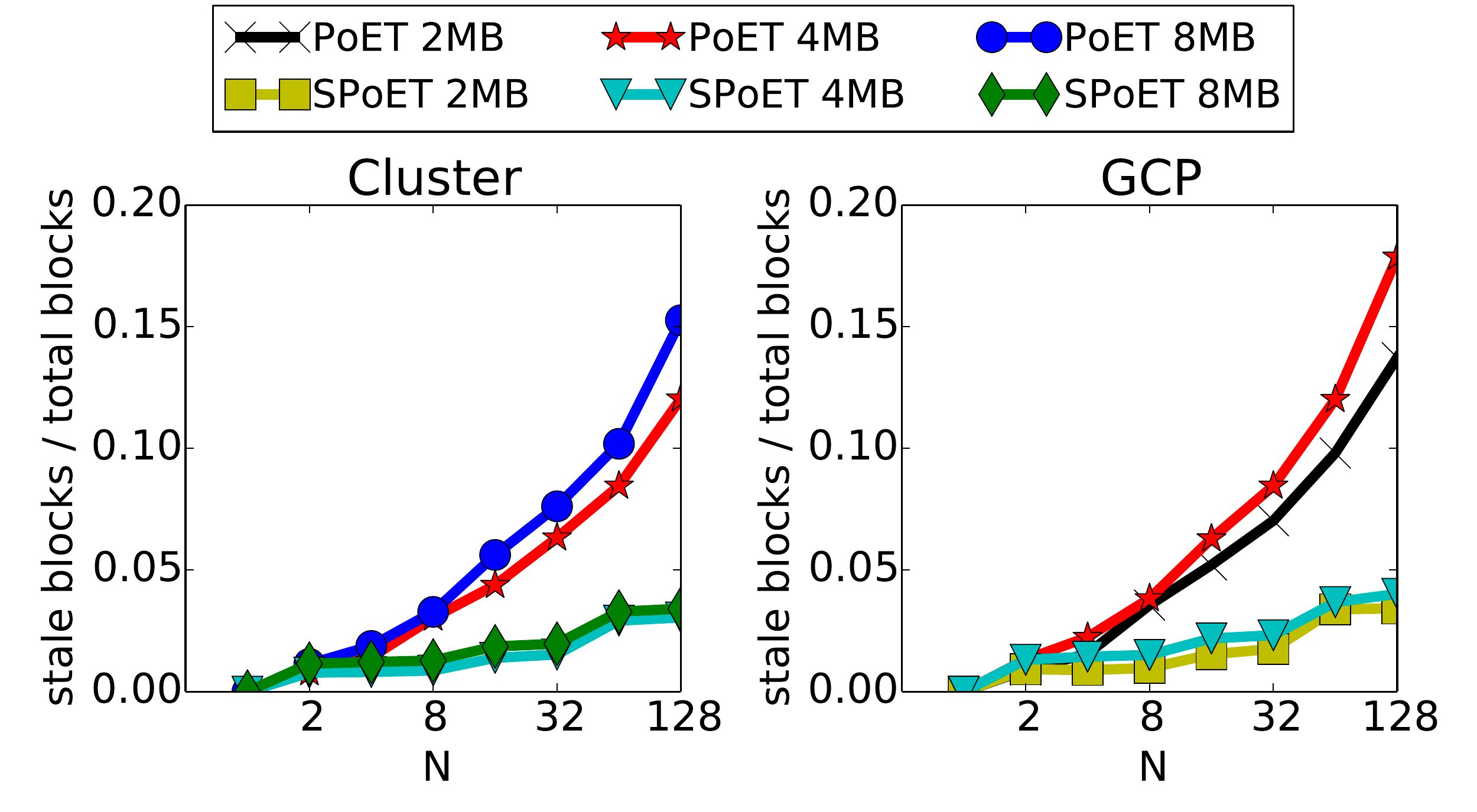}
\caption{\poet\ and \spoet's stale block rate.}
\label{fig:spoet_stale_block}
\end{figure}

The block size is varied from 2MB to 8MB, and the block time from $12s$ to
$24s$.
As $N$ increases, block propagation time, which depends on block time and block
size, increases and leads to higher stale block rate and lower throughput, as
shown in Figure~\ref{fig:spoet} and \ref{fig:spoet_stale_block}.
\spoet\ maintains up to $4\times$ higher throughput because it reduces the stale
block rate significantly from $15\%$ to $3\%$ with $N=128$.

\subsection{Comparison of BFT protocols}
\label{appendix_subsec:bfts}
Figure~\ref{fig:cluster_bft} compares the performance of four popular BFT
implementations in blockchains, namely PBFT, Tendermint, IBFT and Raft (both
from Quorum). For this comparison, we used the key-value benchmark in
BLOCKBENCH. For Tendermint, we used the provided {\tt tm-bench} tool that
benchmarks a simple key-value application.

We examined throughputs with varying fault tolerance, as function of number of
nodes, $N$, and varying workload, as function of number of concurrent clients.
In both settings, PBFT outperforms the alternatives, except at $N=1$ where PBFT
has lower throughput than Tendermint because Hyperledger limits REST request
rate at around $400$ requests per second. These results are due to the design of
Tendermint and IBFT, and due to the implementation of Raft in Quorum.

Unlike PBFT that relies on a stable leader to drive consensus, Tendermint and
IBFT rotate leaders in a round-robin fashion. In particular, nodes in Tendermint
and IBFT may take many rounds, with a new leader at every round, before agreeing
on a block. A node votes to change to a new round when its timer expires. Safety
is achieved via a locking mechanism in which a node locks on a block after it
receives more than $2f+1$ Prepare messages. Once locked, it does not vote for
other blocks. For liveness, the lock is released when there are more than $2f+1$
Prepare messages for another block in a later round. However, we observe that
IBFT suffers from deadlock, because its locks are not released properly. In
Tendermint, a new block can only be proposed when the previous one is finalized
because of locking and rotating the leader. This lock-step execution of
consensus is different from Hyperledger's PBFT where a leader pipelines many
blocks before the first block is finalized. In particular, a node can vote on
many blocks at the same time, if the blocks are assigned consecutive sequence
numbers. Such pipelined execution extracts more concurrency from the consensus
implementation, thereby achieving higher throughput.

Raft has a higher fault tolerance threshold than PBFT ($\frac{N}{2}$ vs.
$\frac{N}{3}$), therefore it is expected to have higher throughput. But we
observed a lower performance for a Raft-based blockchain (Quorum), than a PBFT
blockchain (Hyperledger). This is due to the naive integration of Raft into
Quorum, which fails to pipeline consensus execution. More specifically, a node
in Quorum first constructs a block, then it runs Raft with other nodes to
finalize the block. Next, it constructs a new block and repeats these steps.
Because a block is constructed once the previous block is finalized, consensus
happens in lockstep and the overall throughput suffers as a result.

The difference between Tendermint and Quorum is primarily due to blockchain
features other than consensus.
Tendermint benchmark uses a key-value application that simply stores data tuples
in memory, without any other blockchain features such as Merkle trees and smart
contract execution.
In contrast, a transaction in Quorum is expensive because of its execution in
the EVM and updates to various Merkle trees.

\end{document}